\documentclass[aps,reprint,superscriptaddress,longbibliography]{revtex4-1}

\usepackage[utf8]{inputenc}
\usepackage[dvipsnames]{xcolor}
\usepackage{tikz}
\usetikzlibrary{positioning}\usepackage{setspace}
\usepackage[english]{babel}
\usepackage{lineno}
\usepackage{hyperref}
\usepackage{outlines}
\usepackage{fancyvrb}
\usepackage{amsmath}
\usepackage{hyperref}
\usepackage{braket}
\usepackage{comment}
\usepackage{bbold}
\usepackage{amssymb}
\usepackage{graphicx}
\usepackage[caption=false]{subfig}
\usepackage{url}
\usepackage{soul}
\usepackage{qcircuit}

\def\>{\rangle}
\def\<{\langle}

\newcommand{\map}[1]{\mathcal{#1}}

\usepackage{environ}
\makeatletter
\newsavebox{\measure@tikzpicture}
\NewEnviron{scaletikzpicturetowidth}[1]{%
  \def\tikz@width{#1}%
  \def\tikzscale{1}\begin{lrbox}{\measure@tikzpicture}%
  \BODY
  \end{lrbox}%
  \pgfmathparse{#1/\wd\measure@tikzpicture}%
  \edef\tikzscale{\pgfmathresult}%
  \BODY
}
\makeatother

\begin{document}


\title{Quantum advantage in training binary neural networks}


\author{Yidong Liao}
\affiliation{Institute for Quantum Science and Engineering, Department of Physics, Southern University of Science and Technology (SUSTech),  Shenzhen, China}

\author{Daniel Ebler }
\affiliation{Institute for Quantum Science and Engineering, Department of Physics, Southern  University of Science and Technology (SUSTech),  Shenzhen, China}
\affiliation{Wolfson College, University of Oxford, Linton Road, Oxford OX2 6UD, UK}

\author{Feiyang Liu}
\affiliation{Institute for Quantum Science and Engineering, Department of Physics, Southern  University of Science and Technology (SUSTech),  Shenzhen, China}

\author{Oscar Dahlsten}
\thanks{Correspondence: Oscar Dahlsten \\(dahlsten@sustech.edu.cn)}
\affiliation{Institute for Quantum Science and Engineering, Department of Physics, Southern  University of Science and Technology (SUSTech),  Shenzhen, China}
\affiliation{Center for Quantum Computing, Peng Cheng Laboratory, Shenzhen, 518000, China}
\affiliation{London Institute for Mathematical Sciences, 35a South Street Mayfair, London W1K 2XF, UK}
\affiliation{Wolfson College, University of Oxford, Linton Road, Oxford OX2 6UD, UK}




\begin{abstract}
The performance of a neural network for a given task is largely determined by the initial calibration of the network parameters. Yet, it has been shown that the calibration, also referred to as training, is generally NP-complete. 
This includes networks with binary weights, an important class of networks due to their practical hardware implementations. We therefore suggest an alternative approach to training binary neural networks. It utilizes a quantum superposition of weight configurations. We show that the quantum training guarantees with high probability convergence towards the globally optimal set of network parameters. This resolves two prominent issues of classical training: (1) the vanishing gradient problem and (2) common convergence to suboptimal network parameters. Moreover we achieve a provable polynomial---sometimes exponential---speedup over classical training for certain classes of tasks. We design an explicit training algorithm and implement it in numerical simulations. 

\end{abstract}


\maketitle


\section*{Introduction}
Artificial Neural Networks (NNs) ~\cite{grossberg1988neural,haykin2009neural,russell2016artificial}, have proven immensely successful for a large variety of tasks, notably including pattern recognition~\cite{bishop1995neural, nasrabadi2007pattern}, language processing \cite{collobert2008unified,mikolov2010recurrent}, and simulation of molecular dynamics \cite{kozuch2018combined,degiacomi2019coupling}, leading to changes in practice in fields such as medicine, pharmaceutical research \cite{baxt1991use,karabatak2009expert,zupan1999neural,agatonovic2000basic}, and finance \cite{trippi1992neural}.  While computers were originally built to process information according to a pre-defined algorithm, NNs can learn how to process data themselves. Usually, this is achieved by training the network: data, for which the desired output is already known, is input and the network parameters are adjusted until the outputs of the NN coincide with the desired ones. However, training NNs was shown to be NP-complete \cite{rojas2013neural}, leading to long training times \cite{blum1989training} and large consumption of memory \cite{rumelhart1985learning,werbos1990backpropagation}.

Recently, simplified models such as binary neural networks (BNNs) \cite{hubara2016binarized}, for which the network parameters can only assume bit values, were introduced. Such simplifications drastically reduce the memory size and run time during the execution of the network \cite{Rastegari:tn}. However, during the calibration of parameters, the parameters and outputs of the BNN are kept real and the binarization happens \textit{after} each training cycle \cite{hubara2016binarized}. Hence, even for strongly simplified models improved training methods remain to be found. 

Furthermore, common training methods, such as gradient descent, heavily rely on the shape of the optimisation landscape of the network parameters. Since the landscape is usually not convex, common training methods find suboptimal choices of parameters, corresponding to local extrema rather than global ones \cite{choromanska2015open}. It is known that there are scenarios where local extrema lead to performances which are far from globally optimal \cite{SwirszczCP16,Alizadeh:2018ul}.

A novel approach to NNs is based on quantum technology, which has been shown to achieve performances beyond the possible of current classical implementations for a variety of tasks \cite{deutsch1992rapid,grover1996fast, shor1999polynomial,bravyi2017quantum}. These so-called quantum neural networks (QNNs) strive not only for an improved learning capacity \cite{schuld2014quest}, but also for learning \cite{wan2018learning, morales2018variationally} and identifying new \textit{quantum} protocols \cite{wan2017quantum, beer2019efficient}. An appealing possibility is that quantum effects, such as superposition and entanglement, can also improve the efficiency of training methods. Proposals of quantum neural networks are commonly accompanied by a fully classical \cite{wan2017quantum, bergholm2018pennylane,daskin2018simple,zhao2018forecast} or semi-classical training method assisted with quantum effects \cite{wan2018learning,Farhi:wv,beer2019efficient}, leading to the performance issues discussed above \cite{McClean:2018um}. There have been recent proposals for showing a quantum speed-up for training \cite{verdon2018universal}, taking advantage of a quantum speed-up for gradient descent~\cite{GilyenASW17, Jordan05}. Such results suggest quantum neural networks constitute a well-defined and potentially advantageous route. 
However, a key challenge with gradient descent based training methods is the risk of arriving in local extrema of the optimization landscape. In \cite{ricks2004training} the high-level idea to apply quantum search, which yields optimal solutions, to NN training was suggested, though without a concrete training scheme. There are certain hurdles to directly applying the standard quantum search techniques to neural nets, in particular how to relate the training to a superposition of weight strings that takes into account the cost function and is amenable to a quantum search.

In this work we explicitly propose a method for training binary neural networks using quantum search techniques. The protocol initialises the system parameters coherently and maintains the quantum superposition throughout the full training. By reformulating the training as a quantum search problem, we (1) are guaranteed (with high probability) to find the globally optimal set of network parameters and (2) achieve a provable speed-up in the number of possible choices of parameters, as detailed later in the paper.

We proceed as follows. After a brief introduction to BNNs and Grover's search algorithm, we design a quantum extension of a binary neuron. Then, the fully quantum training protocol is presented. Finally, an analysis of the performance in comparison to classical binary neurons is given and a generalisation from single neurons to feedforward quantum BNNs suggested. The Appendix includes simulations of the quantum training for concrete examples of networks.

\section{Preliminaries}
We commence by introducing classical NNs and their basic training method. Afterwards, we give an overview of Grover's quantum search algorithm and quantum phase estimation.

\subsection{Classical feedforward binary neural networks}
\begin{figure}[h]
\begin{center}
\begin{scaletikzpicturetowidth}{0.5\textwidth}
\begin{tikzpicture}[xscale=0.9, yscale=1.1]

\node (in1) at (-1.5,0) {$a_1$};
\node (in2) at (-1.5,-1.5) {$a_2$};
\node[draw=black,fill=white] (a1) at (0,0) {$\mathcal{N}_1$};
\node[draw=black,fill=white] (b1) at (3,0.5) {$\mathcal{N}_3$};
\node[draw=black,fill=white] (c1) at (6,0) {$\mathcal{N}_6$};
\node[draw=black,fill=white] (a2) at (0,-1.5) {$\mathcal{N}_2$};
\node[draw=black,fill=white] (b2) at (3,-0.75) {$\mathcal{N}_4$};
\node[draw=black,fill=white] (b3) at (3,-2) {$\mathcal{N}_5$};
\node[draw=black,fill=white] (c2) at (6,-1.5) {$\mathcal{N}_7$};
\node (out1) at (7.5,0) {$a'_1$};
\node (out2) at (7.5,-1.5) {$a'_2$};

\draw[->] (in1) -- (a1);
\draw[->] (in2) -- (a2);

\draw[->] (a1) -- node[above]{$w_1$} (b1);
\draw[->] (a1) -- node[below]{$w_2$} (b2);
\draw[->] (a1) -- node[above]{$w_3$} (b3);
\draw[->] (a2) -- node[below=0.5mm]{$w_4$} (b2);
\draw[->] (a2) -- node[above]{$w_5$} (b1);
\draw[->] (a2) -- node[below]{$w_6$} (b3);

\draw[->] (b1) -- node[above]{$w_7$} (c1);
\draw[->] (b1) -- node[above]{$w_8$} (c2);
\draw[->] (b2) -- node[above]{$w_9$} (c2);
\draw[->] (b2) -- node[below=0.6mm]{$w_{10}$} (c1);
\draw[->] (b3) -- node[below]{$w_{11}$} (c1);
\draw[->] (b3) -- node[below]{$w_{12}$} (c2);

\draw[->] (c1) -- (out1);
\draw[->] (c2) -- (out2);

\end{tikzpicture}

\end{scaletikzpicturetowidth}
\end{center}
\caption{\textit{Simple instance of an FNN:} the inputs $a_1$, $a_2$ are forwarded to the neurons $\map N_1$ and $\map N_2$, which process them and forward their outputs to the next layer, consisting of $\map N_3$, $\map N_4$ and $\map N_5$. Afterwards, the outputs of the second layer is forwarded to $\map N_6$ and $\map N_7$, which output $a'_1$, $a'_2$. Each of the three layers is connected with edges carrying weights $w \in \{ w_i \}_{i=1}^{12}$.}
\label{fig:classical_FNN}
\end{figure}
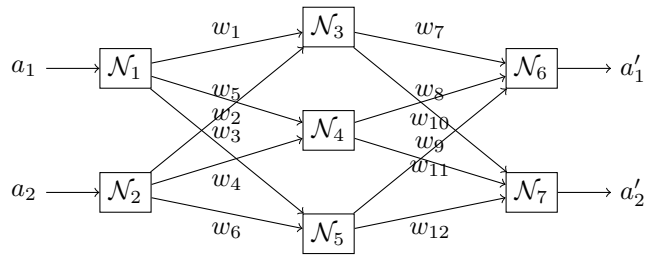

Artificial neural networks (NNs) constitute a computational framework used to process data without pre-defining an algorithm. The task-specific algorithm is constructed from the input data alone, which allows NNs to operate universally, up to restrictions due to the architecture of the network. The basic building block of an NN, the neuron, acts as a computational cell by receiving a set of inputs and processing them according to a pre-set rule. The output is then forwarded to the connected neurons. Modelling the structure as a graph, the edge between two neurons carries a weight $w$, which describes the importance of the connection, see e.~g.~Figure \ref{fig:classical_FNN}.

We focus in this Letter on feedforward NNs (FNNs), for which the neurons are grouped into sequential layers and data is fed in one direction only. Every neuron in layer $r$ shares an edge to every neuron in layer $r-1$ and $r+1$, while the neurons within a layer or across neighbouring layers do not interact (Figure \ref{fig:classical_FNN} depicts a simple instance of an FNN).  The first layer obtains a set of inputs, processes it and forwards it to the next layer. The cascade of computations terminates when the last layer yields the final outputs. Note, that the amount of neurons can be different in each layer.  

The ability of NNs to learn certain tasks like pattern recognition \cite{bishop1995neural,nasrabadi2007pattern} and classification \cite{kotsiantis2006machine} comes with a high computational cost. One of the reasons for this is that the inputs to a neuron and weights $w$ can be arbitrary real numbers, which generally require large amounts of memory and arithmetic operations. Only recently, Bengio \textit{et al} \cite{hubara2016binarized} proposed a simplified version of NNs, which limits the outputs of the neurons and the weights to be binary. As a consequence, the memory size is drastically reduced and most arithmetic operations are replaced with bit-wise operations.
The so-called binary neural networks (BNNs) were shown to process data faster and with a smaller amount of memory compared to NNs with continuous parameters. Remarkably, despite the strong nature of the simplifications the accuracy of BNNs has been shown to be similar to NNs \cite{hubara2016binarized}.
\subsubsection*{Training an NN: gradient descent}
During the training phase of an NN, pairs of training data $(a,a^*)$ consisting of inputs $a$ and corresponding desired outputs $a^*$ (also called labels) are used to calibrate the weights $w$. This happens by sending the inputs through the NN, and comparing the corresponding outputs $a'$ to the correct outputs $a^*$. The weights $w$ in the NN are refined according to the deviation of the output from the desired ones. To quantify the deviation, a task-specific cost function $C(\{a',a^*\})$ is defined. Note, that the outputs $a'$ implicitly depend on the choice of weights. A widely used approach in training classical neural networks is to approach the minimum cost through the method of gradient descent, which updates the weight $w_{ij}$ between neuron $i$ and neuron $j$ from the $t$-th iteration of the NN to the next one as
\begin{align}
w_{ij}(t+1)=w_{ij}(t) - \eta \frac{\partial C}{\partial w_{ij}}  \ . \label{gradientdesc}
\end{align}
Here, $\eta$ is a positive number, usually referred to as the learning rate.

The direct evaluation of the term $\partial C / \partial w_{ij}$ is computationally expensive, and reuquires to loop over all inputs twice. In addition, even if the values of the network parameters and inputs is restricted (e.g. for binary neural networks, see Section \ref{bnnsec}), commonly the weights are kept real-valued and only discretised after the evaluation of the partial derivatives \cite{hubara2016binarized,courbariaux2015binaryconnect}.

A technique known as backpropagation removes the need to evaluate Eq.~(\ref{gradientdesc}) for all weights: given the structure of the NN one can infer how to change weights in earlier layers given Eq.~(\ref{gradientdesc}) for later layers.

Gradient descent, whether with or without back-propagation, has the disadvantage of finding minima that may be local, i.e. sub-optimal \cite{choromanska2015open}. To find a global minimum is in general NP-Hard consistent with the exponential - in the number of neurons -number of weight configurations \cite{rojas2013neural,livni2014computational}. 
\subsection{Grover's search algorithm} 
For an unsorted list of length $N$, identifying the index of a given element in the list requires $O(N)$ computational steps classically. If quantum effects are allowed for, only $O(\sqrt{N})$ steps are needed. This is the result of the famous search algorithm proposed by Grover \cite{grover1996fast}, in which a speedup is achieved by evolving a superposition of list elements. In the following, we give a brief overview of the algorithm.

Let us encode the $N$ items of a list into quantum states in the computational basis $|0\>,|1\>,\dots,|N-1\>$. Furthermore, let $\{|\omega_i\>\}_{i=1}^M \subset \{  |0\>,|1\>,\dots,|N-1\> \}$ be the set of states corresponding to the solutions to the search problem. The algorithm then proceeds as follows:

\begin{itemize}
\item Initialize the registers in the superposition
\begin{align*}
\{  |0\>,|1\>,\dots,|N-1\> \} \mapsto |X\> = \frac{1}{\sqrt{N}}\sum_{x=0}^{N-1} |x\>
\end{align*}
\item For $O(\sqrt{N/M})$ times, repeat:
\begin{enumerate}
\item Apply the quantum oracle $\Lambda_{\omega}=-2 \sum_{i=1}^M |\omega_i\>\<\omega_i| + I$
\item Apply the diffusion transform $D=H^{\otimes n} \Lambda_0 H^{\otimes n}=H^{\otimes n} \left(2 (|0\>\<0|)^{\otimes n} - I \right) H^{\otimes n} $
\end{enumerate}
\item Apply a measurement in the computational basis to the register
\end{itemize}
The quantum oracle $\Lambda_{\omega}$ marks the state $|x\>$ by flipping its sign if it coincides with one of the solutions from the set $\{ \omega_i \}_{i=1}^M$, and does nothing otherwise. The diffusion operator amplifies the amplitude of the marked elements, such that the concluding measurement yields the outcome $x^* \in \{ \omega_i \}_{i=1}^M$ with high probability. 

The optimal number of iterations of the oracle and the diffusion before the concluding measurement is given by $k^*\approx \sqrt{\frac{N}{M}} \frac{\pi}{4}$ \cite{grover1996fast}. It is, hence, crucial to have knowledge about $M$, as stopping at the wrong time may result in a random output rather than a valid solution.

\subsection{Phase Estimation \label{phaseest}}
\begin{figure}[h]
\includegraphics[width=1\linewidth]{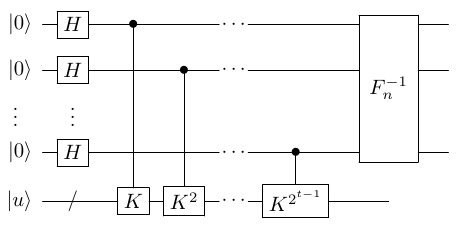}
\caption{Circuit for the quantum Phase estimation: the $t$ ancillary qubits act as control systems for the unitary $K$. The system input $|u\>$ is the corresponding eigenvector  to the eigenvalue $e^{i \pi 2\phi}$ of $K$. The $l$-th ancilla applies $K$ for $2^{l-1}$ times on the system. After the in total $2^t-1$ controlled applications of $K$, the control systems are transformed by an inverse quantum Fourier transform $F_n^{-1}$, which yields a binary encoding of the number $\phi$.}
\label{fig:PE}
\end{figure}

Suppose $|u\>$ is the eigenvector of a unitary operator $K$, with eigenvalue $e^{i \pi 2\phi}$. To estimate $\phi$ up to some error $\epsilon$, we initiate $t$ qubits in the state $|0\>$. The number $t$ is determined by the precision of the estimation of $\phi$, as well as the probability of success of the estimation. The phase estimation algorithm succeeds in two steps and is depicted in Figure \ref{fig:PE}.

\textit{Step 1.} Apply a Hadamard gate to each of the $t$ qubits in the first register, yielding the states $|+\>^{\otimes t}$. Then, the first ancilla acts as a control for $2^0=1$ use of the unitary operator $K$, which is applied to $|u\>$. The second ancilla acts as a control for $2^1=2$ uses of $K$, and so on. In general, the $l$-th ancilla acts as a coherent control for $2^{l-1}$ uses of $K$. In the end, the collective output state $|T\>$ of the $t$ ancillas yields \cite{nielsen2002quantum}
\begin{align*}
|T\>=\frac{1}{\sqrt{2^t}}\sum_{k=0}^{2^t-1} e^{2 \pi i \phi k} |k\> \ .
\end{align*}

\textit{Step 2.} Apply the inverse quantum Fourier Transform $F_n^{-1}$ in order to convert $\phi$ of the state $|T\>$ into a $t$-qubit register. More precisely, the inverse QFT acts as 
\begin{align}
F_n^{-1} |T\> = |\phi_1 \phi_2 \dots \phi_t\> \ , \label{PEout}
\end{align}
where the phase to be estimated is assumed to have the binary form $\phi=0. \phi_1 \phi_2 \dots$, i.e. $\phi_j$ is the $j$-th decimal bit of the number $\phi$. It is important to remark that with the phase estimation (PE) algorithm being unitary, it acts linearly on superpositions $\sum_i |u_i\>$ of eigenvectors $|u_i\>$ \cite{nielsen2002quantum}.

\section{Quantum binary neurons \label{bnnsec}}

A classical neuron is the basic computational unit of a classical NN. It implements a function, which takes multiple inputs and produces a single output. Hence, this many-to-one mapping makes the classical neuron function irreversible, which is a priori not compatible with reversible dynamics of quantum computing. However, this mapping can be generalised to a quantum neuron in the same way that quantum computing generalises classical computing, along the paradigm of \cite{wan2017quantum}. To define the quantum generalisation, first, classical gates are extended to reversible gates, which are a subset of quantum unitaries. Afterwards, general unitaries are allowed for.

\subsection{Finding a reversible extension}
\begin{figure}[h]
\begin{center}
\begin{scaletikzpicturetowidth}{0.49\textwidth}
\begin{tikzpicture}[scale=\tikzscale]
\draw (0.8,-1.5) node {$w_1$};
\draw (0.8,-1.9) node {$a_1$}; 
\draw (0.8,-2.5) node {$w_2$};
\draw (0.8,-2.9) node {$a_2$};

\draw (1.0,-1.5) -- (1.25,-1.5);
\draw (1.0,-1.9) -- (1.25,-1.9);
\draw (1.0,-2.5) -- (1.25,-2.5);
\draw (1.0,-2.9) -- (1.25,-2.9);

\draw (1.25,-1.3) rectangle (2.5,-2.1);
\draw (1.85,-1.7) node {$\tt XNOR$};

\draw (1.25,-2.3) rectangle (2.5,-3.1);
\draw (1.85,-2.7) node {$\tt XNOR$};

\draw (2.5,-1.7) -- (2.75,-1.7);
\draw (2.5,-2.7) -- (2.75,-2.7);

\draw (3,-1.7) node {$s_1$};
\draw (3,-2.7) node {$s_2$};

\draw (3.25,-1.7) -- (3.5,-1.7);
\draw (3.25,-2.7) -- (3.5,-2.7);

\draw (3.5,-1.5) rectangle (4.75,-2.9);
\draw (4.1,-2) node {$\tt bit-$};
\draw (4.15,-2.4) node {$\tt count$};

\draw (4.75,-2.3) -- (5,-2.3);
\draw (5.25,-2.3) node {$s$};
\draw (5.5,-2.3) -- (5.75,-2.3);

\draw[dashed] (5.75,-1.5) rectangle (7,-2.9);
\draw (5.9,-2.7) -- (6.3,-2.7) -- (6.3,-1.7) -- (6.8,-1.7);
\draw[->] (7,-2.3) -- (7.5,-2.3);
\draw (8.6,-2.3) node {$f(s) = a'$};
\end{tikzpicture}
\end{scaletikzpicturetowidth}
\end{center}
\caption{\textit{Functioning of a CBN:} an $\tt XNOR$ operation multiplies the inputs $a_1$, $a_2$ with the corresponding weights $w_1$, $w_2$. The resulting values $s_1=w_1 a_1$, $s_2=w_2 a_2$ are forwarded to the $\tt bitcount$, which outputs $s=s_1+s_2$. Finally, the Sign function determines the activation of the neuron.}
\label{fig:CBN}
\end{figure}
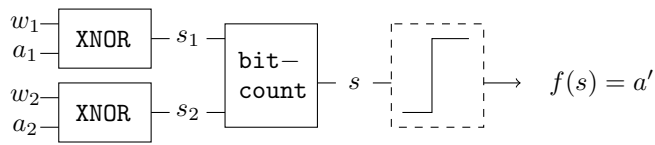

Let us consider the elementary instance of a classical binary neuron (CBN) depicted in Figure \ref{fig:CBN}. 

\begin{figure}[h]
\begin{center}
\begin{scaletikzpicturetowidth}{0.3\textwidth}
\begin{tikzpicture}[scale=\tikzscale]
\draw (0.8,-4) node {$w_1$};
\draw (0.8,-4.4) node {$a_1$}; 
\draw (0.8,-5) node {$w_2$};
\draw (0.8,-5.4) node {$a_2$};

\draw (1.0,-4) -- (1.25,-4);
\draw (1.0,-4.4) -- (1.25,-4.4);
\draw (1.0,-5) -- (1.25,-5);
\draw (1.0,-5.4) -- (1.25,-5.4);

\draw (1.25,-3.8) rectangle (2.5,-4.6);
\draw (1.85,-4.2) node {$U_{\tt XNOR}$};

\draw (1.25,-4.8) rectangle (2.5,-5.6);
\draw (1.85,-5.2) node {$U_{\tt XNOR}$};
 
\draw (2.5,-4) -- (2.75,-4);
\draw (2.5,-4.4) -- (2.75,-4.4);
\draw (2.5,-5) -- (2.75,-5);
\draw (2.5,-5.4) -- (2.75,-5.4);

\draw (3,-4.4) node {$s_1$};
\draw (3,-5.4) node {$s_2$};
\draw (3,-5.8) node {$0$};

\draw (3.25,-4.4) -- (3.5,-4.4);
\draw (3.25,-5.4) -- (3.5,-5.4);
\draw (3.25,-5.8) -- (3.5,-5.8);

\draw (3.5,-4.2) rectangle (4.75,-6);

\draw (4.1,-5.1) node {$U_{\tt bit+}$};

\draw (4.75,-4.4) -- (5,-4.4);
\draw (4.75,-5.4) -- (5,-5.4);
\draw (4.75,-5.8) -- (5,-5.8);

\draw (5.25,-5.8) node {$a'$};
\end{tikzpicture}
\end{scaletikzpicturetowidth}
\end{center}
\caption{\textit{Reversible embedding of CBN:} the gate $U_{\tt XNOR}$ take component $i$ of the input vector $\underline{a}$ and multiplies it with the corresponding edge weight $w_i$, yielding the value $s_i=a_i w_i$, with $i\in \{ 1,2 \}$. To achieve reversibility $w_i$ is forwarded on the first output. The gate $U_{\tt bit+}$ takes $s_1$, $s_2$ and an ancilla $0$, and encodes the activation $f(s)$ in the output $a'$, while preserving $s_1$ and $s_2$.}
\label{revCBN}
\end{figure}
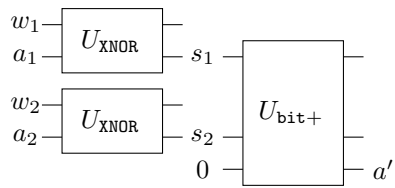

The neuron takes the input $\underline{a}=(a_1,a_2)$  with the two components $a_1$ and $a_2$, and has weights $w_1,w_2 \in \{1,-1 \}$ on the edges. An $\tt XNOR$ gate multiplies the inputs with the weights, yielding $s_i=w_i a_i$, $i=1,2$. Afterwards, the operation $\tt bitcount$ sums the results to the value $s=w_1 a_1 + w_2 a_2$. Finally, the Sign function $f(s)$ determines the activation value $a'$ of the neuron. Now, firstly we introduce ancillary inputs and outputs for each operation in the CBN, such that the number of inputs coincides with the number of outputs. These ancillas carry the output values of the operations, while the inputs are preserved. This yields a reversible embedding $U_{\tt XNOR}$ of $\tt XNOR$ and $U_{\tt bit+}$ of $\tt bitcount$, see Figure \ref{revCBN}. Here, $U_{\tt bit+}$ includes both the $\tt bitcount$ operation and the activation through the activation function $f$.

\subsection{Unitary embedding of the reversible CBN \label{basicopqbn}}
The reversible logical gates in Figure \ref{revCBN}, with inputs $x_i$ and outputs $y_i$ can be implemented as quantum unitaries $U=\sum_i |y_i\>\<x_i|$. This defines a quantum binary neuron (QBN). In order to find a quantum circuit implementation of a QBN, we decompose the operations $U_{\tt XNOR}$ and $U_{\tt bit+}$ into elementary quantum gates.

\begin{figure}[h]
\begin{center}
\begin{scaletikzpicturetowidth}{0.3\textwidth}
\begin{tikzpicture}[scale=\tikzscale]
\draw (0.7,-6.5) node {$\ket{w_1}$};
\draw (0.68,-6.9) node {$\ket{a_1}$}; 
\draw (0.7,-7.5) node {$\ket{w_2}$};
\draw (0.7,-7.9) node {$\ket{a_2}$};

\draw (1.0,-6.5) -- (1.25,-6.5);
\draw (1.0,-6.9) -- (1.25,-6.9);
\draw (1.0,-7.5) -- (1.25,-7.5);
\draw (1.0,-7.9) -- (1.25,-7.9);

\draw (1.25,-6.3) rectangle (2.5,-7.1);
\draw (1.85,-6.7) node {$\text{CNOT}$};
\draw (1.25,-7.3) rectangle (2.5,-8.1);
\draw (1.85,-7.7) node {$\text{CNOT}$};
 
\draw (2.5,-6.5) -- (2.75,-6.5);
\draw (2.5,-6.9) -- (2.75,-6.9);
\draw (2.5,-7.5) -- (2.75,-7.5);
\draw (2.5,-7.9) -- (2.75,-7.9);

\draw (3,-6.9) node {$\ket{s_1}$};
\draw (3,-7.9) node {$\ket{s_2}$};
\draw (3,-8.3) node {$\ket{0}$};

\draw (3.25,-6.9) -- (3.5,-6.9);
\draw (3.25,-7.9) -- (3.5,-7.9);
\draw (3.25,-8.3) -- (3.5,-8.3);

\draw (3.5,-6.7) rectangle (4.75,-8.5);

\draw (4.1,-7.6) node {$U_{\tt bit+}$};

\draw (4.75,-6.9) -- (5,-6.9);
\draw (4.75,-7.9) -- (5,-7.9);
\draw (4.75,-8.3) -- (5,-8.3);

\draw (5.25,-8.3) node {$\ket{a'}$};
\end{tikzpicture}
\end{scaletikzpicturetowidth}
\end{center}
\caption{\textit{Functioning of QBN:} the input data is encoded into quantum states. The multiplication succeeds by the CNOT gate and outputs the states $|s_1\>$, $|s_2\>$. Finally, the Toffoli gate, which has known decompositions into elementary gates, executes $U_{\tt bit+}$, leading to the output state $|a'\>$. }
\label{fig:QBN}
\end{figure}
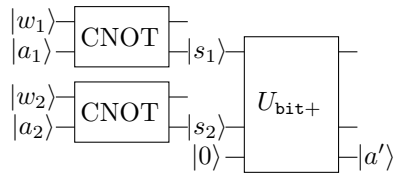

The multiplication of the input and weight can be achieved by a controlled NOT (CNOT) gate on each input $|a_i \>$, controlled by the state of the corresponding weight $|w_i\>$ (see Figure \ref{fig:QBN}).

It can be easily seen from the truth table of the $\tt XNOR$ and CNOT gate that the operations indeed coincide. The gate $U_{\tt bit+}$ is realized by the Toffoli gate \cite{barenco1995elementary} on the ancilla, where the states $|s_1\>$ and $|s_2\>$ act as control systems.

\subsection{Generalization to quantum FBNNs \label{networkgeneralisation}}
So far, we proposed a unitary extension of a classical neuron. A neural network consists of multiple interlinked neurons. Hence, by extending the neurons unitarily, the resulting  quantum neural network acts, overall, unitarily on inputs as well. Generalising Figure \ref{fig:QBN}, a quantum feedforward binary neural network (QFBNN) takes a set of $N$ weights $\{ w_1, w_2, \dots, w_N\}$ as input, together with a set of data $\underline{a}=(a_1,a_2,\dots,a_p)$. The set $\{ w_1, w_2, \dots, w_N\}$ denotes all weights in the networks -- including input layer, hidden layers and output layer. The set of data $\underline{a}=(a_1,a_2,\dots,a_p)$ assumes that there are $p$ neurons in the input layer (with assigned weights $w_1,w_2,\dots,w_p$). Each of the $r$ neurons in the last layer of the QFBNN outputs one value, leading to the overall output $\underline{a}'=(a_1',a_2',\dots a_r')$. Then, by depicting the action of the QBFNN as a single unitary $U$, the fully quantum training method introduced in the next Section generalises straightforwardly from a single neuron to general QFBNNs.

One point requires a bit more attention: when generalising from a single neuron to a feedforward neural network, the output values of the neurons are to be copied and distributed to each neuron in the next layer. While classically this is trivial to do, quantum mechanics prohibits exact copying of data -- a phenomenon called no-cloning theorem \cite{wootters1982single}. For classical binary neural networks, this issue is resolved by applying CNOT gates to the output of each neuron, one for each copy operation. The CNOT acts on an ancillary qubit initialized in $|0\>$, controlled by the output qubit $|a'\>$  of the neuron. Hence, for bit valued $a'$ the CNOT gate acts as
\begin{align}
\text{CNOT} (|a'\> |0\>) = |a'\>|a'\> \ .
\end{align}
In total from one layer with $\ell_1$ neurons to the next with $\ell_2$ neurons, we need $\ell_1 \times  \ell_2$ ancillas. When the overall state has coherence, the CNOT gates in general do not produce perfect copies but rather entanglement between the states and the ancillas carrying the output values.

\section{Quantum training with phase estimation \label{petrain}}
Here, we propose a fully quantum training protocol for QBNs, meaning that we initialise the weights of the network in a quantum superposition and do not collapse the state throughout the training. For presentational clarity we restrict the training to a single neuron. This readily generalises to networks using the recipe described in section~\ref{networkgeneralisation}.  The general strategy proceeds in the following three steps

\begin{enumerate}
\item Each weight string in superposition is multiplied by a phase factor which depends on how many times it leads to the correct output, for the input data in the training set. This is achieved by looping over training data coherently, making use of a technique known as uncomputation.
\item Binarisation of the marking: the phase factors from step 1 are binarised to the values $\{1, -1\}$ using quantum phase estimation.
\item Step 1 and 2 together yield an oracle compatible with Grover search which is applied as a final step.
\end{enumerate}
This yields with high probability the globally optimal set of weights, quadratically - in the number of weight strings - faster than a brute force classical search. While the main text discusses the subroutines in detail, Appendix \ref{app:qbnnexp} illustrates the functioning of the training method with concrete examples.
\subsection{Marking the target weights \label{markingtheweights}} 
Let us assume we have in total $N$ binary weights $\{ w_1, w_2, \dots , w_N \}$, with values encoded as $|0\>$ and $|1\>$. Consequently, there are in total $\tilde{N}=2^N$ different combinations of values for the weights. Each of these combinations will be denoted by a string $\underline{w}$ of length $N$, for which the corresponding quantum state reads $|\underline{w}\>$. Furthermore, let $n$ be the number of data pairs $(\underline{a_1},a^*_1),(\underline{a_2},a^*_2),\dots,(\underline{a_n},a^*_n)$ used for the training. Here, the value $a^*_i$ (sometimes called label) notes the desired output of the given input $\underline{a_i}$. 

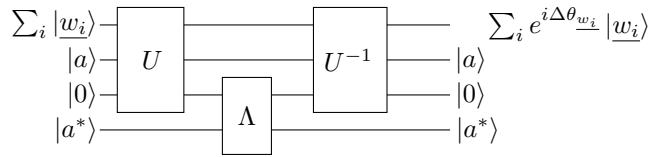
\begin{figure}[h]
\begin{center}
\begin{scaletikzpicturetowidth}{0.48\textwidth}
\begin{tikzpicture}[scale=\tikzscale]
\draw (0,1.5) -- (5,1.5);
\draw (0,2) -- (5,2);
\draw (0,2.5) -- (5,2.5);
\draw (0,3) -- (5,3);

\draw (-0.65,3) node {$\sum_i \ket{\underline{w_i}}$};
\draw (-0.25,2.5) node {$\ket{a}$};
\draw (-0.25,2) node {$\ket0$};
\draw (-0.34,1.5) node {$\ket{a^*}$};

\draw[fill=white] (0.25,3.25) rectangle (1.2,1.75);
\draw (0.75,2.5) node {$U$};

\draw[fill=white] (1.75,2.25) rectangle (2.45,1.15);
\draw (2.1,1.7) node {$\Lambda$};

\draw[fill=white] (3.05,3.25) rectangle (4.1,1.75);
\draw (3.55,2.5) node {$U^{-1}$};

\draw (6.7,3) node {$\sum_i e^{i \Delta \theta_{\underline{w_i}}}\ket{\underline{w_i}}$};
\draw (5.3,2.5) node {$\ket{a}$};
\draw (5.3,2) node {$\ket0$};
\draw (5.4,1.5) node {$\ket{a^*}$};
\end{tikzpicture}
\end{scaletikzpicturetowidth}
\end{center}
\caption{\textit{Marking of the target weights:} the QBN of Figure \ref{fig:QBN} acts as a unitary $U$ on the input $|a\>$, ancilla $|0\>$ and coherent weight state $|W\>$. The output $|a'\>$ is compared to the desired output $|a^*\>$ by the oracle $\Lambda$. To decouple the weights, an uncomputation $U^{-1}$ is applied. The weights are now in state $|W'\>=\sum_i e^{i \Delta \theta_{\underline{w_i}}}\ket{\underline{w_i}}$.}
\label{fig:marking}
\end{figure}

The first subroutine works as a phase accumulation approach to identify the goodness of a string, see Figure \ref{fig:marking}. The resulting phases will be used later to mark the good strings for a quantum search.

\textit{Step 1:} initialize all $\tilde{N}=2^N$ possible weight strings in a coherent superposition $|W\>=\frac{1}{\sqrt{\tilde{N}}} \sum_i |\underline{w_i}\>$. This yields the overall initial state 
\begin{align}
|\text{In}_1\>= |W\> |\underline{a}\> |a^*\> |0\> \ ,
\end{align}
where $|0\>$ is an ancillary qubit and $(\underline{a},a^*)\in \{(\underline{a_1},a^*_1), (\underline{a_2},a^*_2),\dots,(\underline{a_n},a^*_n) \}$ is one pair of training data. Due to the superposition of weight states, all weight states are simultaneously combined with the input $|\underline{a}\>$. Next, the unitary action $U$ (see Figure \ref{fig:QBN}) of the QBN acts on $|W\>$, the input $|\underline{a}\>$ and an ancilla $|0\>$, encoding the output $|a'\>$ on the ancillary system. The overall state is then given by
\begin{align}
|\text{Out}_1\>=U |\text{In}_1\>= \frac{1}{\sqrt{\tilde{N}}} \sum_i |\underline{w_i}, \underline{\widetilde{a_i}} ,a_i'\> |a^*\> \ , \label{markingout1}
\end{align}
where the QBN transforms the input $|\underline{a}\>$ and ancilla into $|\underline{\widetilde{a_i}}\>$ and  $|a_i'\>$ respectively if the control state is $|\underline{w_i}\>$. Note that except for desired out $|a^*\>$ the systems in Eq.~(\ref{markingout1}) are entangled in general. \\

\textit{Step 2:} call the oracle $\Lambda$ to compare the output $|a_i'\>$ with the desired output $|a^*\>$. If the two states coincide  $\Lambda$ adds a phase $e^{i \pi/n}$ to $|a_i'\>$. This leads to

\begin{align}
\label{eq:step2}
|\text{Out}_2\>= \frac{1}{\sqrt{\tilde{N}}} \sum_i |\underline{w_i}, \underline{\widetilde{a_i}}, a_i'\> e^{i  \Delta \theta_{\underline{w_i}} } |a^*\>  \ ,
\end{align}

where $\Delta \theta_{\underline{w_i}} = \pi/n$  if the oracle comparison between $a_i'$ and $a^*$ was successful, and $\Delta \theta_{\underline{w_i}}=0$ otherwise. \\

\textit{Step 3:} decouple the weights by uncomputation. By inverting the unitary action $U$ of the QBN, the weights get decoupled and assume the state
\begin{align}
|W'\>=\frac{1}{\sqrt{\tilde{N}}} \sum_i e^{i \Delta \theta_{\underline{w_i}}} |\underline{w_i}\> \ .
\end{align}
The full output state is given by $|\text{Out}_3\> = |W'\> |\underline{a}\> |a^*\> |0\>$.
\newline 

\textit{Step 4:} accumulation of phases. The state $|W'\>$ is used as the initial weight state for the marking with a new training pair. By repeating Step 1 - 3 for all $n$ pairs of training data, we achieve a coherent phase accumulation for all weight strings, resulting in the output weight state
\begin{align}
|\widetilde{W}\>=\frac{1}{\sqrt{\tilde{N}}} \sum_i e^{i \pi N_i /n} |\underline{w_i}\> \ . \label{phaseaccweights}
\end{align} 
Here, $N_i \leq n$ denotes the number of times the oracle comparison was successful for weight string $\underline{w_i}$ during the $n$ rounds. The most frequently marked weights are closest to a phase of -1, whereas bad weight strings maintain a phase of $+1$. In Appendix ~\ref{app:qbnnexp} we discuss the quantum circuit implementations of several basic examples of the above marking scheme and their simulation results. The simulations have been done via Huawei's quantum computing cloud platform HiQ(\url{https://hiq.huaweicloud.com}), as numerical evidence of the proposed quantum training algorithm. \\

Now, let $\tilde{U}$ be the unitary describing the full accumulation process, namely 
\begin{align*}
&\tilde{U} \left(|W \> \otimes \left( \bigotimes_{j=1}^n  |\underline{a_j}\> \otimes |a^*_j\> \right) \otimes |\underline{0}\> \right) \nonumber \\
=& |\widetilde{W} \> \otimes \left( \bigotimes_{j=1}^n  |\underline{a_j}\> \otimes |a^*_j\> \right) \otimes |\underline{0}\> \ ,
\end{align*}
where $|\underline{0}\>= |0\>^{\otimes n}$. It can be seen easily that the states $|\underline{w_i}\> \otimes ( \bigotimes_{j=1}^n |\underline{a_j}\> \otimes |a^*_j\>) \otimes  |\underline{0}\>$, $i=1,2,\dots,n$ are eigenvectors of $\tilde{U}$ with corresponding eigenvalues $e^{i \pi N_i /n}$. This is an important insight for next step in which the phases will be converted to a binarized marking via phase estimation.

\subsection{Binarizing the marking}
The output state of Eq. \ref{phaseaccweights} suggests to use a generalisation of the standard marking of Grover search to oracles which add certain phases rather than $+1$ or $-1$ to the elements subject to the search. It was shown in \cite{grover1998quantum} that this is indeed possible, maintaining the scaling $O(\sqrt{N/M})$ of the number of iterations. However, the optimal number $k^*$ of iterations of the algorithm is in general unknown which makes such a generalisation impractical.

With the aim of making the output amenable to Grover search, we thus propose a subroutine to binarize the phases in the coherent weight state $|\tilde{W}\>$. To this purpose, we define a threshold count $N_t$, which denotes the minimum required number of successful oracle calls for a weight string to get marked. If $N_i \geq N_t$, the corresponding weight state will get a phase of $-1$. If $N_i < N_t$, the phase is set to 1. This conversion can be done via phase estimation as shown in the following Section.

\subsubsection{Binarised marking via phase estimation}
\begin{figure}[h]
\includegraphics[width=1\linewidth]{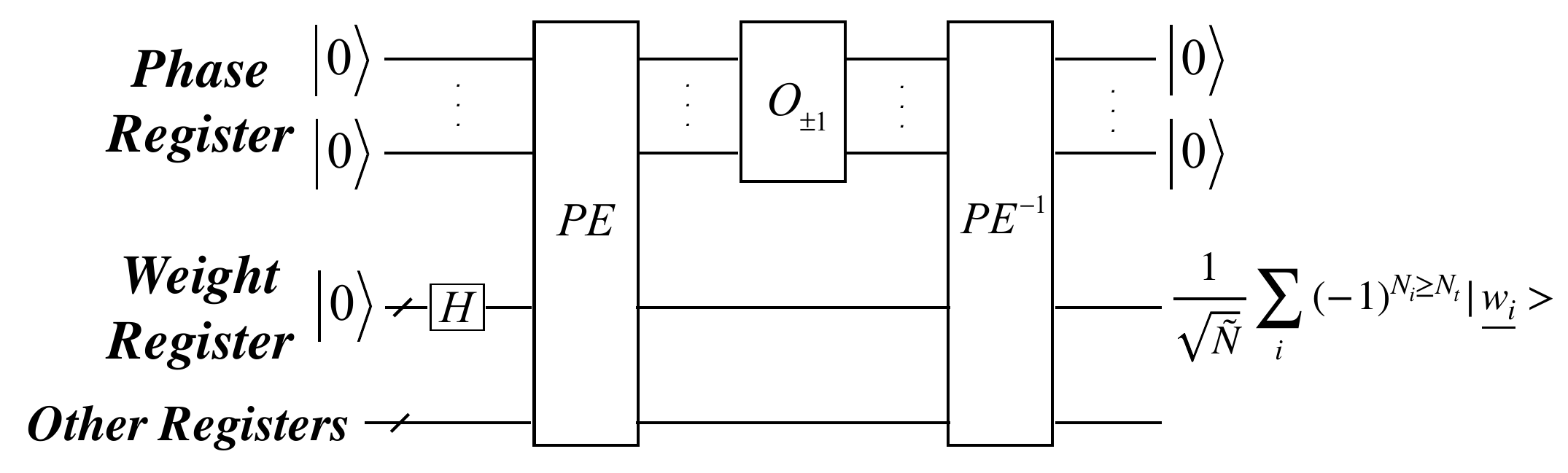}
\caption{\textit{Binarised marking via phase estimation:} In the first step, the PE is applied to the coherent superposition $|W\>$ of weight states, together with all other inputs and ancillas included in the training. The unitary action $K$ is the full marking scheme $\tilde{U}$ described in Section \ref{markingtheweights}. This yields the binary encoding $|\phi^{(i)}_1 \phi^{(i)}_2 \dots \phi^{(i)}_t\>$ of the phases $\phi^{(i)}=N_i/2n$ subject to the estimation. Then, the oracle $O_{\pm 1}$ is called, which will add a marker to the state if the phase is larger than a certain threshold $N_t$. Finally, inverting the PE achieves a decoupling of the marked weight state from the other systems.}
\label{fig:PEfull}
\end{figure}

The complete process of converting the phases in Eq. \ref{phaseaccweights} to a binary marking is visualized in Figure \ref{fig:PEfull}.

For our training algorithm, the eigenvectors are given by $|u_i\>=|\underline{w_i}\> \otimes \bigotimes_{j=1}^n (|\underline{a_j}\> \otimes |a^*_j\>) \otimes  |\underline{0}\>$. Then, we identify the controlled unitary operation $K$ acting on the superposition (we omit normalisation for readability) $|u\> = \sum_{i=1}^{\tilde{N}} |u_i\>$ as the full phase accumulation in Section \ref{markingtheweights}, i.e. $K=\tilde{U}$, and the numbers to be estimated $\phi^{(i)} = N_i / (2 n)$. We now define a new oracle $O_{\pm 1}$, acting as
\begin{align*}
O_{\pm 1} |\phi_1 \phi_2 \dots \phi_t \> = \pm 1 |\phi_1 \phi_2 \dots \phi_t \> \ ,
\end{align*}
where $O_{\pm 1}$ adds $-1$ if $N_i \geq N_t$. In general, we can make certain choices for the threshold $N_t$, such that $N_t/2n$ corresponds to a binary fraction of length $\ell$. Then, for $t$ large enough it can be determined from the estimate $|\phi_1 \phi_2 \dots \phi_t \>$ if $N_i \geq N_t$ is met.

For illustration purposes, let us analyse some simple examples. Assume first we have $t=1$, such that PE gives the first decimal bit $\phi_1$. Then, using $\phi = N_i / (2 n)$, we have $\phi_1=1$ if $N_i=n$, and $\phi_1=0$ if $N_i< n$. If $t=2$ PE gives $\phi_1 \phi_2$, which is $\phi_1 \phi_2= 0 1$ if $N_i \geq n/2$ and $\phi_1 \phi_2= 00$ if $N_i \leq n/2$. For $t=3$, PE yields $\phi_1 \phi_2 \phi_3$, which assumes the value $011$ if $N_i \geq 3 n/4$, and $010$ if $n/2 \leq N_i < 3 n/4 $. Thus, even though the number of uses of $\tilde{U}$ grows exponentially in $t$, a small number of $t$ is sufficient for a good precision in singling out the best weight strings. Note that $t$ only depends on our required precision, not on $N$.

We comment in Section \ref{bsearch} in greater detail how to choose $t$ and in Section \ref{perfPE} its impact on the performance of the marking.

Finally, after PE and the oracle $O_{\pm 1}$, the weight state together with the phase register state reads
\begin{align}
|\tilde{\tilde{W}}\>=\frac{1}{\sqrt{\tilde{N}}}  \sum_i (-1)^{N_i \geq N_t}  |\underline{w_i}\> |\phi^{(i)}_1 \phi^{(i)}_2 \dots \phi^{(i)}_t \> \ , \label{weightunc}
\end{align}
where the exponent $N_i \geq N_t$ is to be understood as a Boolean variable which is 1 in case the condition is true and 0 else. 

The state in Eq. (\ref{weightunc}) is entangled between the weight strings and the binary encoding of phases from the PE. In order to disentangle the two systems while maintaining coherence of the weight state, we uncompute the PE, which transforms (\ref{weightunc}) to
\begin{align}
PE^{-1}: |\tilde{\tilde{W}}\> \mapsto |\hat{W}\>=  \frac{1}{\sqrt{\tilde{N}}}  \sum_i (-1)^{N_i \geq N_t} |\underline{w_i}\> \ . \label{weightfin}
\end{align}
In the next step, the amplitude of the marked strings in the state $|\hat{W}\>$ are amplified such that the concluding measurement detects one of the marked strings with high probability.

\subsection{Full training cycle \label{fulltrain}}
\begin{figure}[h]
\includegraphics[width=1.00\linewidth]{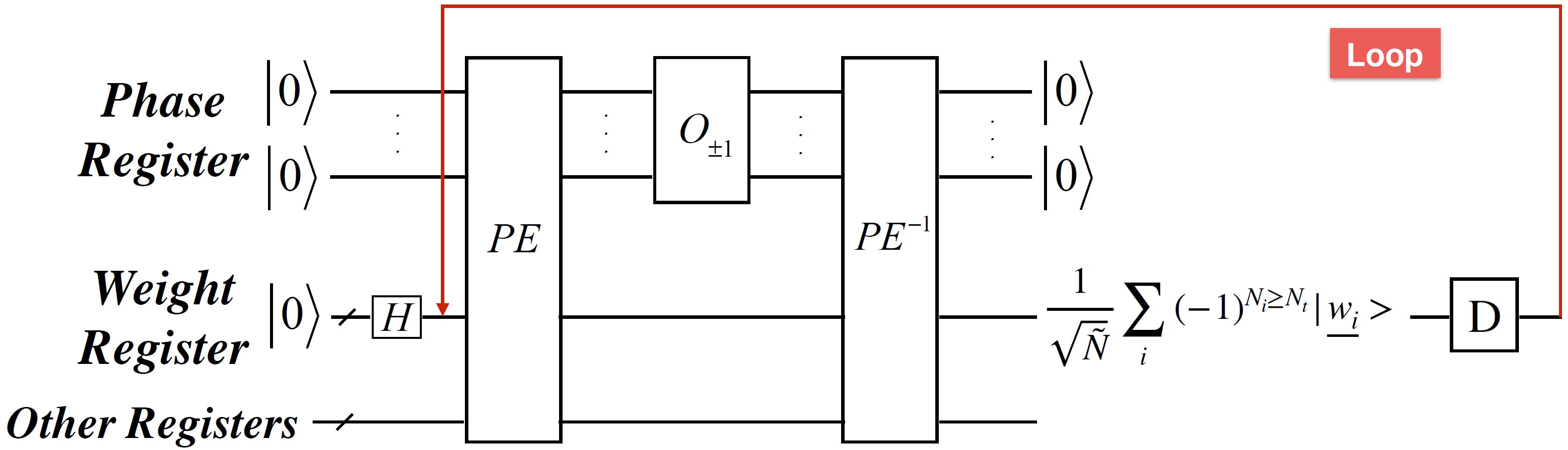}

\caption{\textit{Full traning cycle with phase estimation:} After the binary marking process depicted in Figure \ref{fig:PEfull}, the diffusion operator $D=H \Lambda_{0} H$ is applied to the weight register. Amplification of the marked elements is achieved by looping the full marking subroutine and the subsequent diffusion for $k^*$ times. A final measurement in the computational basis yields the optimal weight string $\underline{w^*}$.}

\label{fig:cycle}
\end{figure}

It can be seen from Eq.~(\ref{weightfin}) that the marked state $|\hat{W}\>$ is equivalent to the marked state of standard Grover search. Hence, in order to amplify the amplitudes of the marked elements the standard diffusion operator $D=H^{\otimes n} \Lambda_{0} H^{\otimes n}$ can be used. The quantum training for the binary neuron then succeeds by applying $k^*\approx \sqrt{\frac{\tilde{N}}{M}} \frac{\pi}{4}$ iterations of the binary marking via phase estimation and the subsequent diffusion, if we assume that we know the number $M$ of target weight strings. If $M$ is not known one can find  an optimal weight string in an expected time that scales as $O(\sqrt{\tilde{N}/M})$ with a subroutine based on Boyer \textit{et al.} \cite{boyer1998tight}. Figure \ref{fig:cycle} summarizes the full training cycle with phase estimation. A circuit implementation of the full training is given in Appendix \ref{app:fulltrain}.

\subsubsection*{Binary search for globally optimal solution \label{bsearch}}
For a fixed precision $t$ of the phase estimation subroutine, the number of solutions $M$ depends on the training set and is generally unknown. Indeed, there might be multiple marked weight strings, or even none which is above the threshold $N_t$. To resolve this problem, we propose a binary search on the threshold intervals as a subroutine to ensure, up to the precision of the search and up to imprecisions in the phase estimation algorithm, that the optimal solution is marked and amplified. 

For a solution $w^*$ the possible range of the number of matches $N^*$ in the accumulative marking scheme is initially an integer from the interval $[0,n]$. At every step of the search the possible range is cut evenly into two parts. Thus the size of the possible range goes as $n2^{-i}$ for the i-th step of the binary search. The search stops when the size of the possible range reaches a small enough number which quantifies our desired precision, and which we call $\delta < n$, i.e. the search runs until i is large enough so that $n2^{-i}\leq \delta$. To determine whether the updated possible range is the lower or upper half, we mark the weight strings in the upper half by -1 using the procedures described above. If after the search the measured weight string is in the upper half this becomes the new possible range. Else if it is not in the upper half the possible range becomes the lower half. Thus eventually this binary search identifies 
with a high probability an $N^*$ which is within $\delta$ of the global optimum. This takes  $\lceil \log (\frac{n}{\delta})\rceil$ steps, which scales efficiently in the number of $n$. \\

The binary search then proceeds as follows: 

\begin{itemize}
\item Initialize $N_t = n/2$, $i=0$. \\Set $m=1$ and pick $\mu \in (1,4/3)$.
\item For $i \leq \lceil \log(n/\delta) \rceil$, loop the following steps: 
\begin{enumerate}
\item Choose an integer $s\in[1,m-1]$ uniformly at random.
\item With current $N_t$, apply $s$ iterations of the full training cycle from Figure \ref{fig:cycle}. Measure the weight state and obtain an candidate solution $\underline{w}^*$.
\item Run the classical BNN on $\underline{w}^*$ only to obtain $N^*$, the number of times $\underline{w}^*$ gave correct outputs. 

\item \textit{If} $N^* \geq N_t $, this means the observed $\underline{w}^*$ is indeed a solution for current $N_t$, and there might be better solutions for larger $N_t$, so update $N_t$ as:
$N_t \rightarrow N_t + \Delta N_i /2$, in which $\Delta N_i=n2^{-i}$ , and go to next step to update i. \\

\textit{else} (i.e. $N^* < N_t $), set $m$ to $\min(\mu m, \sqrt N)$, and go back to step 1. \\
\textit{Time-out} rule for this inner loop on $m$: when the loop of step 1-4 continues to the case that the full training cycle has been totally executed $O(\sqrt N)$ iterations, but $N^* < N_t $ still, this means there is no solution for the current $N_t$, then update $N_t$ as:
$N_t \rightarrow N_t - \Delta N_i /2$ in which $\Delta N_i=n2^{-i}$ , and go to next step to update i.

\item update i as: $i \rightarrow i+1$.
\end{enumerate}
\item output the last value of $\underline{w}^*$ that has $N^* \geq N_t $ and end training. This $\underline{w}^*$ is the optimal weights the algorithm found.

\end{itemize} 

The above search homes in on a solution that is globally optimal up to the $\delta$ precision.
The number of steps in the loop (on i) of the binary search can by inspection be seen to be scale as $\log(\frac{n}{\delta})$.

\section{Quantum training with counter registers}
Quantum training via phase estimation includes an overhead  of neuron calls that is exponential in the precision $t$ of the phase estimation. There is an alternative approach which is sensitive to the size of the training set, but avoids that overhead: rather than accumulating phases, the number of matches between the QBN output and the desired output can be encoded directly into a multi-qubit register.

The algorithm is similar to the training cycle presented in Section \ref{petrain}, up to some changes: in Figure \ref{fig:marking} the oracle $\Lambda$ adding incremental phases to good weight strings is replaced by a new oracle $O_{\tt count}$, acting on a $\lceil \log(n+1) \rceil$-qubit register as 
\begin{align}
O_{\tt count} |h\> = \begin{cases} |h+1\> \ \text{if} \ a'=a^* \\
 |h\> \ \text{else}  \end{cases} \ .
\end{align}
The number $h$ of matches is then encoded in binary into the qubits. Initially the counter is set to $|0\>$ and increased coherently during the marking. This way, after the full set of training data, the output weight state reads
\begin{align}
|\tilde{W}\>=\frac{1}{\tilde{N}} \sum_i  |\underline{w_i}\> |N_i\> \ ,
\end{align}
where $N_i$ is the amount of matches for the weight string $\underline{w_i}$.

Next, we again define a threshold count $N_t$ and a new oracle $\widetilde{O}_{\pm 1}$. If the condition $N_i \geq N_t$ holds, the oracle marks the corresponding weight string $\underline{w_i}$ by adding a factor of $-1$. Otherwise, the string is left invariant. The resulting state reads
\begin{align}
|\tilde{\tilde{W}}\>=\frac{1}{\sqrt{\tilde{N}}} \sum_i (-1)^{N_i \geq N_t} |\underline{w_i} \> |N_i\> \ .
\end{align}
In order to retrieve the same marked state as in standard Grover search, the counter needs to be reset. To maintain coherence, such a reset needs to be unitary. This can be achieved by  uncomputing the $n$ rounds of counting at the beginning of the marking. 



 The weight state then becomes 
\begin{align}
|\hat{W}\>=\frac{1}{\sqrt{\tilde{N}}} \sum_i (-1)^{N_i \geq N_t} |\underline{w_i} \>  \ ,
\end{align}
and the same binary search procedure can be applied.

\section{Performance \label{perfPE}}
One way to quantify the time requirement for the training is to count (1) the number of calls to the neural network and (2) the number of calls to the oracles. Recall there are two more elementary oracles, one for comparing the output with the desired output, and one for checking if phase is above a threshold. These together with the phase accumulation routine acts as a Grover Oracle. Counting the calls to the oracles as a way of quantifying performance is reasoned since the probability of a system being lost, or the amount of noise more generally, will often scale as the number of calls. Moreover, the cost to experimenters is often the amount of time the experiment takes, which may also scale as the number of calls. 

In the quantum training presented above there are two separate contributions to the number of calls: (i) for each training input $(\underline{a_i},a_i^*)$ there is one call to the accumulation subroutine (consisting of one call to the comparatory oracle and two neural network calls: one call and one inverse call for the uncomputation), and there are $n$ accumulation cycles in total (recall $n$ is the number of training inputs), (ii) the quantum search protocol on the weight states is expected to take a number of calls scaling as the $O \left(\sqrt{\tilde{N}/M} \right)$, with $M$ being the estimated number of solutions and $\tilde{N}=2^N$ being the total number of weight states. 

Hence, in terms of the number of training data pairs, $n$, and the total number of weight states $\tilde{N}$ , we call the combined Grover oracle 
\begin{align}
\sum_{i=1}^{\lceil \log{n/ \delta} \rceil} \sqrt{\frac{\tilde{N}}{M_i}}\leq \lceil \log\left(\frac{n}{\delta}\right) \rceil \sqrt{\tilde{N}} \equiv N_G \label{groveroraclecalls}
\end{align}
times, where  $M_i$ is the number of optimal weights for the $i$-th loop in the binary search and $M_i \geq 1$. The sum in Equation \ref{groveroraclecalls} comes from the binary search with precision $\delta$ described in Section \ref{fulltrain}. The comparatory oracle is then called 
\begin{align}
N_C=N_G \left(2 n \left(\sum_{j=0}^{t-1} 2^j \right) \right) = N_G \left(2 n \left(2^t - 1 \right)\right)  \approx 4 N_G n^2 
\label{groveroraclecalls2}
\end{align}
times (the factor of 2 comes from the uncomputation of the phase estimation), and the QBNN is called $2 N_C$ times. The last approximation follows from the fact that the estimated phases are given by $N_i/2n$ and the smallest possible precision of $N_i$ is $\delta=1$. In order to be able to give an estimate up to $\delta=1$, we require the $t$-th binary digit in $0.\phi^{(i)}_1 \phi^{(i)}_2 \dots \phi^{(i)}_t$ to be such that $2^{-t}=1/2n$. This gives $2^t - 1 \approx 2n$.\\

The number of qubits needed constitutes of (i) the number of qubits needed for the phase estimation and (ii) the number of qubits needed for QBNN. 

For (i), the phase estimation requires $\log (2n/\delta)$ ancillary qubits to estimate with precision $\delta$ of the binary search, regardless of the imprecisions in the phase estimation algorithm itself. For (ii), the number of qubits needed for the QBNN is consists of the amount of qubits needed for the training inputs, the desired outputs, the $N$ weight qubits, and the ancilla qubits. There are two types of ancillas: the ones to achieve the unitary embedding of the classical binary neural network and the ones needed for the fan-out operation. We show in Appendix \ref{app:noofqubits} that the following following simple equality holds
\begin{align}
Q_{\text{input}} + Q_{\text{ancilla}} = Q_{\text{weight}} + Q_{\text{output}} \ .
\end{align}
Here, $Q_{\text{input}}$ is the amount of input qubits, $Q_{\text{ancilla}}$ the amount of ancilla qubits, $Q_{\text{weight}}$ the amount of weight qubits and $Q_{\text{output}}$ the number of output qubits. 
It can be easily seen that the amount of qubits needed for the desired outputs is the same as $Q_{\text{output}}$.
Therefore, the total number of qubits needed for the QBNN is equal to
\begin{align}
 2 (Q_{\text{weight}} + Q_{\text{output}} ) \leq 2 (Q_{\text{weight}} + Q_{\text{input}})
\end{align}

\subsection{Classical vs. quantum training}
The training of classical neural networks are known to have three drawbacks: (1) there is no efficient method, and indeed training has been shown to be NP-complete even for small networks \cite{blum1989training,rojas2013neural}. (2) Commonly, the solutions found in training methods based on gradient descent are only locally optimal and can be globally very suboptimal \cite{SwirszczCP16,Alizadeh:2018ul}. (3) The gradient of the cost function with respect to a specific weight can be vanishingly small, preventing convergence to a solution, or explodingly large. This issue is often referred to as the problem of vanishing or exploding gradients. 

In the following we elaborate on how the quantum training method in Section \ref{fulltrain} addresses these drawbacks.
 \subsubsection{Quantum training faster} 
The outcome of the quantum training methods discussed above is guaranteed with high probability to yield the globally optimal set of weights. For sufficiently unstructured cost landscapes, training methods based on gradient descent commonly output weight configurations corresponding to local extrema of the cost function. As a result, the performance on fresh data after the training stage risk being poor. 

To ensure classically that the globally optimal parameters are found would require a brute-force search over a list of size $n \times \widetilde{N}$,  in which $\tilde{N} = 2^N$ and $N$ is the number of weights in the network, $n$ is the number of training samples in the training set. This list consist of all weight strings and input pairs with assignment of a cost value. As each call of that list requires comparing desired outputs with outputs, such a brute force search calls the comparatory oracle ${N_C}^{classical}=n\times\tilde{N}$ times. 
On the other hand, from Eq.\ref{groveroraclecalls} and Eq.\ref{groveroraclecalls2}, our quantum training calls the comparatory oracle ${N_C}^{quantum} \approx 4n^2\log\left(\frac{n}{\delta}\right)\sqrt{\tilde{N}} $ times, where $\delta$ represents the precision of the threshold in the binary search for globally optimal weights (see section \ref{bsearch}), therefore the ratio of ${N_C}^{classical} $ and  ${N_C}^{quantum}$ is:
\begin{align}
\frac{{N_C}^{classical} }{{N_C}^{quantum} } \approx \frac{ \sqrt{\tilde{N}}}{4n\log\left(\frac{n}{\delta}\right)} = \frac{2^{N/2}}{4n\log\left(\frac{n}{\delta}\right)},
\label{ratio}
\end{align}
Note that, for training to be practical, $n$ should scale reasonably with the number of weights, and is thus expected to be bounded by a polynomial of $N$. Thus Eq.\ref{ratio} can be termed an exponential advantage in $N$ for such cases. For example, within the architecture of a two-hidden-layer feedforward network that are enough to learn $n$ training samples \cite{huang2003learning}, we have the relation $n < N$ (see Appendix \ref{2hidden} for details). Therefore for this example we have:
\begin{equation}
\frac{{N_C}^{classical} }{{N_C}^{quantum} } >\frac{ 2^{N/2}}{4N\log\left(\frac{N}{\delta}\right) }
\label{advantage}
\end{equation}
 
Eq.~\ref{advantage} shows an instance of the quantum advantage of our training protocol over classical brute force search. The value of $N$ is often large, in the order of thousands to millions\cite{mocanu2018scalable}. 
 

\subsubsection{Overcoming vanishing and exploding gradients}
During the classical training, the weights are updated with respect to their caused change in cost. Binary activation functions are usually replaced by continuous approximations such as the sigmoid function. Yet, it can be shown that the gradient of the sigmoid function is bound to the interval $(0,1)$. Hence, in a network with $L$ layers, approximating the derivative $\partial C / \partial w$ of the cost function $C$ with respect to a weight $w$ in layer $\ell$ by backpropagation leads to a multiplication of $\gamma=L-\ell$ terms, all smaller than one -- the gradient vanishes exponentially in the number $\gamma$. Conversely, in certain cases the derivatives can be very large and the derivative explodes. Both situations lead to a failure in training, as the changes in the weights are either vanishingly small or too large to control.

Common approaches to address this issue is using specific activation functions, such as the ReLU function, which improves the issue but does not resolve it in general \cite{glorot2011deep}. Other solutions were proposed, such as classical global search methods \cite{bengio1994learning,Shang485892} and special architectures for neural networks \cite{watrous1992induction,graves2008novel,sutskever2011generating}, but also these approaches were shown to suffer from significant drawbacks \cite{hochreiter1998vanishing} or fail to resolve the exploding gradient problem \cite{pascanu2013difficulty}.

The quantum training method proposed in this work is a global search. It, therefore, finds globally optimal solutions while at the same time avoiding gradient related issues. While classically it is still an active field of research to improve on this problem, the quantum search completely resolves it. This makes the quantum search a powerful option to train classical networks.

\section{Summary and Outlook}
In this work we presented a fully quantum protocol for the training of classical feedforward binary neural networks. It resolves two prominent issues of gradient descent training: suboptimal choices of weights and training failure due to vanishing or exploding gradients.  The protocol is guaranteed to find the globally optimal set of weights, while yielding a quadratic advantage in the number of queries needed to achieve the same with a classical algorithm. A key contribution of the protocol is to find an explicit way of turning the neural net training problem into a quantum search problem. The training protocol is fully specified, and numerical simulations of small but non-trivial quantum binary neural nets were made. 

An advantage with making this quantum generalisation of binary neural nets is that they can be implemented with physical qubits and are thus suitable for realisation with quantum computing experiments. Whilst implementing large networks is currently not feasible, training a single binary neuron--an elementary module of the network--with the method described above requires approximately 8 logical qubits (see the performance section), which is within range of current state-of-the-art experiments.

The ability of finding the globally optimal weights may also induce a risk of overfitting more. There are several methods for avoiding overfitting classically, including dropping out certain layers at times during the training, restricting the number of weights and weight regularisation techniques~\cite{srivastava2014dropout,shrestha2019review}. It thus seems well-motivated to build on this quantum training framework to explore analogous quantum techniques. For example quantum versions of drop-out could involve dynamic quantum network architectures e.g.\ using quantum control systems or teleportation  between non-successive layers. \\

\section*{Acknowledgements}
We would like to thank Yu Hao, Jon Allcock, Joshua Morris, Sina Salek and Christa Zoufal for useful discussions.

\bibliographystyle{apsrev4-1}
\bibliography{main}

\begin{thebibliography}{61}%
\makeatletter
\providecommand \@ifxundefined [1]{%
 \@ifx{#1\undefined}
}%
\providecommand \@ifnum [1]{%
 \ifnum #1\expandafter \@firstoftwo
 \else \expandafter \@secondoftwo
 \fi
}%
\providecommand \@ifx [1]{%
 \ifx #1\expandafter \@firstoftwo
 \else \expandafter \@secondoftwo
 \fi
}%
\providecommand \natexlab [1]{#1}%
\providecommand \enquote  [1]{``#1''}%
\providecommand \bibnamefont  [1]{#1}%
\providecommand \bibfnamefont [1]{#1}%
\providecommand \citenamefont [1]{#1}%
\providecommand \href@noop [0]{\@secondoftwo}%
\providecommand \href [0]{\begingroup \@sanitize@url \@href}%
\providecommand \@href[1]{\@@startlink{#1}\@@href}%
\providecommand \@@href[1]{\endgroup#1\@@endlink}%
\providecommand \@sanitize@url [0]{\catcode `\\12\catcode `\$12\catcode
  `\&12\catcode `\#12\catcode `\^12\catcode `\_12\catcode `\%12\relax}%
\providecommand \@@startlink[1]{}%
\providecommand \@@endlink[0]{}%
\providecommand \url  [0]{\begingroup\@sanitize@url \@url }%
\providecommand \@url [1]{\endgroup\@href {#1}{\urlprefix }}%
\providecommand \urlprefix  [0]{URL }%
\providecommand \Eprint [0]{\href }%
\providecommand \doibase [0]{http://dx.doi.org/}%
\providecommand \selectlanguage [0]{\@gobble}%
\providecommand \bibinfo  [0]{\@secondoftwo}%
\providecommand \bibfield  [0]{\@secondoftwo}%
\providecommand \translation [1]{[#1]}%
\providecommand \BibitemOpen [0]{}%
\providecommand \bibitemStop [0]{}%
\providecommand \bibitemNoStop [0]{.\EOS\space}%
\providecommand \EOS [0]{\spacefactor3000\relax}%
\providecommand \BibitemShut  [1]{\csname bibitem#1\endcsname}%
\let\auto@bib@innerbib\@empty
\bibitem [{\citenamefont {Grossberg}(1988)}]{grossberg1988neural}%
  \BibitemOpen
  \bibfield  {author} {\bibinfo {author} {\bibfnamefont {S.~E.}\ \bibnamefont
  {Grossberg}},\ }\href@noop {} {\emph {\bibinfo {title} {Neural networks and
  natural intelligence}}}\ (\bibinfo  {publisher} {The MIT press},\ \bibinfo
  {year} {1988})\BibitemShut {NoStop}%
\bibitem [{\citenamefont {Haykin}(2009)}]{haykin2009neural}%
  \BibitemOpen
  \bibfield  {author} {\bibinfo {author} {\bibfnamefont {S.~S.}\ \bibnamefont
  {Haykin}},\ }\href@noop {} {\emph {\bibinfo {title} {Neural networks and
  learning machines}}},\ Vol.~\bibinfo {volume} {3}\ (\bibinfo  {publisher}
  {Pearson Upper Saddle River},\ \bibinfo {year} {2009})\BibitemShut {NoStop}%
\bibitem [{\citenamefont {Russell}\ and\ \citenamefont
  {Norvig}(2016)}]{russell2016artificial}%
  \BibitemOpen
  \bibfield  {author} {\bibinfo {author} {\bibfnamefont {S.~J.}\ \bibnamefont
  {Russell}}\ and\ \bibinfo {author} {\bibfnamefont {P.}~\bibnamefont
  {Norvig}},\ }\href@noop {} {\emph {\bibinfo {title} {Artificial intelligence:
  a modern approach}}}\ (\bibinfo  {publisher} {Malaysia; Pearson Education
  Limited,},\ \bibinfo {year} {2016})\BibitemShut {NoStop}%
\bibitem [{\citenamefont {Bishop}\ \emph {et~al.}(1995)\citenamefont {Bishop}
  \emph {et~al.}}]{bishop1995neural}%
  \BibitemOpen
  \bibfield  {author} {\bibinfo {author} {\bibfnamefont {C.~M.}\ \bibnamefont
  {Bishop}} \emph {et~al.},\ }\href@noop {} {\emph {\bibinfo {title} {Neural
  networks for pattern recognition}}}\ (\bibinfo  {publisher} {Oxford
  university press},\ \bibinfo {year} {1995})\BibitemShut {NoStop}%
\bibitem [{\citenamefont {Nasrabadi}(2007)}]{nasrabadi2007pattern}%
  \BibitemOpen
  \bibfield  {author} {\bibinfo {author} {\bibfnamefont {N.~M.}\ \bibnamefont
  {Nasrabadi}},\ }\href@noop {} {\bibfield  {journal} {\bibinfo  {journal}
  {Journal of Electronic Imaging}\ }\textbf {\bibinfo {volume} {16}},\ \bibinfo
  {pages} {049901} (\bibinfo {year} {2007})}\BibitemShut {NoStop}%
\bibitem [{\citenamefont {Collobert}\ and\ \citenamefont
  {Weston}(2008)}]{collobert2008unified}%
  \BibitemOpen
  \bibfield  {author} {\bibinfo {author} {\bibfnamefont {R.}~\bibnamefont
  {Collobert}}\ and\ \bibinfo {author} {\bibfnamefont {J.}~\bibnamefont
  {Weston}},\ }in\ \href@noop {} {\emph {\bibinfo {booktitle} {Proceedings of
  the 25th international conference on Machine learning}}}\ (\bibinfo
  {organization} {ACM},\ \bibinfo {year} {2008})\ pp.\ \bibinfo {pages}
  {160--167}\BibitemShut {NoStop}%
\bibitem [{\citenamefont {Mikolov}\ \emph {et~al.}(2010)\citenamefont
  {Mikolov}, \citenamefont {Karafi{\'a}t}, \citenamefont {Burget},
  \citenamefont {{\v{C}}ernock{\`y}},\ and\ \citenamefont
  {Khudanpur}}]{mikolov2010recurrent}%
  \BibitemOpen
  \bibfield  {author} {\bibinfo {author} {\bibfnamefont {T.}~\bibnamefont
  {Mikolov}}, \bibinfo {author} {\bibfnamefont {M.}~\bibnamefont
  {Karafi{\'a}t}}, \bibinfo {author} {\bibfnamefont {L.}~\bibnamefont
  {Burget}}, \bibinfo {author} {\bibfnamefont {J.}~\bibnamefont
  {{\v{C}}ernock{\`y}}}, \ and\ \bibinfo {author} {\bibfnamefont
  {S.}~\bibnamefont {Khudanpur}},\ }in\ \href@noop {} {\emph {\bibinfo
  {booktitle} {Eleventh Annual Conference of the International Speech
  Communication Association}}}\ (\bibinfo {year} {2010})\BibitemShut {NoStop}%
\bibitem [{\citenamefont {Kozuch}\ \emph {et~al.}(2018)\citenamefont {Kozuch},
  \citenamefont {Stillinger},\ and\ \citenamefont
  {Debenedetti}}]{kozuch2018combined}%
  \BibitemOpen
  \bibfield  {author} {\bibinfo {author} {\bibfnamefont {D.~J.}\ \bibnamefont
  {Kozuch}}, \bibinfo {author} {\bibfnamefont {F.~H.}\ \bibnamefont
  {Stillinger}}, \ and\ \bibinfo {author} {\bibfnamefont {P.~G.}\ \bibnamefont
  {Debenedetti}},\ }\href@noop {} {\bibfield  {journal} {\bibinfo  {journal}
  {Proceedings of the National Academy of Sciences}\ }\textbf {\bibinfo
  {volume} {115}},\ \bibinfo {pages} {13252} (\bibinfo {year}
  {2018})}\BibitemShut {NoStop}%
\bibitem [{\citenamefont {Degiacomi}(2019)}]{degiacomi2019coupling}%
  \BibitemOpen
  \bibfield  {author} {\bibinfo {author} {\bibfnamefont {M.~T.}\ \bibnamefont
  {Degiacomi}},\ }\href@noop {} {\bibfield  {journal} {\bibinfo  {journal}
  {Structure}\ }\textbf {\bibinfo {volume} {27}},\ \bibinfo {pages} {1034}
  (\bibinfo {year} {2019})}\BibitemShut {NoStop}%
\bibitem [{\citenamefont {Baxt}(1991)}]{baxt1991use}%
  \BibitemOpen
  \bibfield  {author} {\bibinfo {author} {\bibfnamefont {W.~G.}\ \bibnamefont
  {Baxt}},\ }\href@noop {} {\bibfield  {journal} {\bibinfo  {journal} {Annals
  of Internal Medicine}\ }\textbf {\bibinfo {volume} {115}},\ \bibinfo {pages}
  {843} (\bibinfo {year} {1991})}\BibitemShut {NoStop}%
\bibitem [{\citenamefont {Karabatak}\ and\ \citenamefont
  {Ince}(2009)}]{karabatak2009expert}%
  \BibitemOpen
  \bibfield  {author} {\bibinfo {author} {\bibfnamefont {M.}~\bibnamefont
  {Karabatak}}\ and\ \bibinfo {author} {\bibfnamefont {M.~C.}\ \bibnamefont
  {Ince}},\ }\href@noop {} {\bibfield  {journal} {\bibinfo  {journal} {Expert
  Systems with Applications}\ }\textbf {\bibinfo {volume} {36}},\ \bibinfo
  {pages} {3465} (\bibinfo {year} {2009})}\BibitemShut {NoStop}%
\bibitem [{\citenamefont {Zupan}\ \emph {et~al.}(1999)\citenamefont {Zupan},
  \citenamefont {Gasteiger},\ and\ \citenamefont {Zupan}}]{zupan1999neural}%
  \BibitemOpen
  \bibfield  {author} {\bibinfo {author} {\bibfnamefont {J.}~\bibnamefont
  {Zupan}}, \bibinfo {author} {\bibfnamefont {J.}~\bibnamefont {Gasteiger}}, \
  and\ \bibinfo {author} {\bibfnamefont {J.}~\bibnamefont {Zupan}},\
  }\href@noop {} {\emph {\bibinfo {title} {Neural Networks in Chemistry and
  Drug Design}}}\ (\bibinfo  {publisher} {Wiley-VCH Weinheim},\ \bibinfo {year}
  {1999})\BibitemShut {NoStop}%
\bibitem [{\citenamefont {Agatonovic-Kustrin}\ and\ \citenamefont
  {Beresford}(2000)}]{agatonovic2000basic}%
  \BibitemOpen
  \bibfield  {author} {\bibinfo {author} {\bibfnamefont {S.}~\bibnamefont
  {Agatonovic-Kustrin}}\ and\ \bibinfo {author} {\bibfnamefont
  {R.}~\bibnamefont {Beresford}},\ }\href@noop {} {\bibfield  {journal}
  {\bibinfo  {journal} {Journal of Pharmaceutical and Biomedical Analysis}\
  }\textbf {\bibinfo {volume} {22}},\ \bibinfo {pages} {717} (\bibinfo {year}
  {2000})}\BibitemShut {NoStop}%
\bibitem [{\citenamefont {Trippi}\ and\ \citenamefont
  {Turban}(1992)}]{trippi1992neural}%
  \BibitemOpen
  \bibfield  {author} {\bibinfo {author} {\bibfnamefont {R.~R.}\ \bibnamefont
  {Trippi}}\ and\ \bibinfo {author} {\bibfnamefont {E.}~\bibnamefont
  {Turban}},\ }\href@noop {} {\emph {\bibinfo {title} {Neural networks in
  finance and investing: Using artificial intelligence to improve real world
  performance}}}\ (\bibinfo  {publisher} {McGraw-Hill, Inc.},\ \bibinfo {year}
  {1992})\BibitemShut {NoStop}%
\bibitem [{\citenamefont {Rojas}(2013)}]{rojas2013neural}%
  \BibitemOpen
  \bibfield  {author} {\bibinfo {author} {\bibfnamefont {R.}~\bibnamefont
  {Rojas}},\ }\href@noop {} {\emph {\bibinfo {title} {Neural networks: a
  systematic introduction}}}\ (\bibinfo  {publisher} {Springer Science \&
  Business Media},\ \bibinfo {year} {2013})\BibitemShut {NoStop}%
\bibitem [{\citenamefont {Blum}\ and\ \citenamefont
  {Rivest}(1989)}]{blum1989training}%
  \BibitemOpen
  \bibfield  {author} {\bibinfo {author} {\bibfnamefont {A.}~\bibnamefont
  {Blum}}\ and\ \bibinfo {author} {\bibfnamefont {R.~L.}\ \bibnamefont
  {Rivest}},\ }in\ \href@noop {} {\emph {\bibinfo {booktitle} {Advances in
  neural information processing systems}}}\ (\bibinfo {year} {1989})\ pp.\
  \bibinfo {pages} {494--501}\BibitemShut {NoStop}%
\bibitem [{\citenamefont {Rumelhart}\ \emph {et~al.}(1985)\citenamefont
  {Rumelhart}, \citenamefont {Hinton},\ and\ \citenamefont
  {Williams}}]{rumelhart1985learning}%
  \BibitemOpen
  \bibfield  {author} {\bibinfo {author} {\bibfnamefont {D.~E.}\ \bibnamefont
  {Rumelhart}}, \bibinfo {author} {\bibfnamefont {G.~E.}\ \bibnamefont
  {Hinton}}, \ and\ \bibinfo {author} {\bibfnamefont {R.~J.}\ \bibnamefont
  {Williams}},\ }\href@noop {} {\emph {\bibinfo {title} {Learning internal
  representations by error propagation}}},\ \bibinfo {type} {Tech. Rep.}\
  (\bibinfo  {institution} {California Univ San Diego La Jolla Inst for
  Cognitive Science},\ \bibinfo {year} {1985})\BibitemShut {NoStop}%
\bibitem [{\citenamefont {Werbos}(1990)}]{werbos1990backpropagation}%
  \BibitemOpen
  \bibfield  {author} {\bibinfo {author} {\bibfnamefont {P.~J.}\ \bibnamefont
  {Werbos}},\ }\href@noop {} {\bibfield  {journal} {\bibinfo  {journal}
  {Proceedings of the IEEE}\ }\textbf {\bibinfo {volume} {78}},\ \bibinfo
  {pages} {1550} (\bibinfo {year} {1990})}\BibitemShut {NoStop}%
\bibitem [{\citenamefont {Hubara}\ \emph {et~al.}(2016)\citenamefont {Hubara},
  \citenamefont {Courbariaux}, \citenamefont {Soudry}, \citenamefont
  {El-Yaniv},\ and\ \citenamefont {Bengio}}]{hubara2016binarized}%
  \BibitemOpen
  \bibfield  {author} {\bibinfo {author} {\bibfnamefont {I.}~\bibnamefont
  {Hubara}}, \bibinfo {author} {\bibfnamefont {M.}~\bibnamefont {Courbariaux}},
  \bibinfo {author} {\bibfnamefont {D.}~\bibnamefont {Soudry}}, \bibinfo
  {author} {\bibfnamefont {R.}~\bibnamefont {El-Yaniv}}, \ and\ \bibinfo
  {author} {\bibfnamefont {Y.}~\bibnamefont {Bengio}},\ }in\ \href@noop {}
  {\emph {\bibinfo {booktitle} {Advances in neural information processing
  systems}}}\ (\bibinfo {year} {2016})\ pp.\ \bibinfo {pages}
  {4107--4115}\BibitemShut {NoStop}%
\bibitem [{\citenamefont {{Rastegari}}\ \emph {et~al.}(2016)\citenamefont
  {{Rastegari}}, \citenamefont {{Ordonez}}, \citenamefont {{Redmon}},\ and\
  \citenamefont {{Farhadi}}}]{Rastegari:tn}%
  \BibitemOpen
  \bibfield  {author} {\bibinfo {author} {\bibfnamefont {M.}~\bibnamefont
  {{Rastegari}}}, \bibinfo {author} {\bibfnamefont {V.}~\bibnamefont
  {{Ordonez}}}, \bibinfo {author} {\bibfnamefont {J.}~\bibnamefont {{Redmon}}},
  \ and\ \bibinfo {author} {\bibfnamefont {A.}~\bibnamefont {{Farhadi}}},\
  }\href@noop {} {\bibfield  {journal} {\bibinfo  {journal} {arXiv e-prints}\
  ,\ \bibinfo {eid} {arXiv:1603.05279}} (\bibinfo {year} {2016})},\ \Eprint
  {http://arxiv.org/abs/1603.05279} {arXiv:1603.05279 [cs.CV]} \BibitemShut
  {NoStop}%
\bibitem [{\citenamefont {Choromanska}\ \emph {et~al.}(2015)\citenamefont
  {Choromanska}, \citenamefont {LeCun},\ and\ \citenamefont
  {Arous}}]{choromanska2015open}%
  \BibitemOpen
  \bibfield  {author} {\bibinfo {author} {\bibfnamefont {A.}~\bibnamefont
  {Choromanska}}, \bibinfo {author} {\bibfnamefont {Y.}~\bibnamefont {LeCun}},
  \ and\ \bibinfo {author} {\bibfnamefont {G.~B.}\ \bibnamefont {Arous}},\ }in\
  \href@noop {} {\emph {\bibinfo {booktitle} {Conference on Learning Theory}}}\
  (\bibinfo {year} {2015})\ pp.\ \bibinfo {pages} {1756--1760}\BibitemShut
  {NoStop}%
\bibitem [{\citenamefont {Swirszcz}\ \emph {et~al.}(2016)\citenamefont
  {Swirszcz}, \citenamefont {Czarnecki},\ and\ \citenamefont
  {Pascanu}}]{SwirszczCP16}%
  \BibitemOpen
  \bibfield  {author} {\bibinfo {author} {\bibfnamefont {G.}~\bibnamefont
  {Swirszcz}}, \bibinfo {author} {\bibfnamefont {W.~M.}\ \bibnamefont
  {Czarnecki}}, \ and\ \bibinfo {author} {\bibfnamefont {R.}~\bibnamefont
  {Pascanu}},\ }\href@noop {} {\bibfield  {journal} {\bibinfo  {journal}
  {arXiv:1611.06310 [stat.ML]}\ } (\bibinfo {year} {2016})}\BibitemShut
  {NoStop}%
\bibitem [{\citenamefont {Alizadeh}\ \emph {et~al.}(2019)\citenamefont
  {Alizadeh}, \citenamefont {Fernández-Marqués}, \citenamefont {Lane},\ and\
  \citenamefont {Gal}}]{Alizadeh:2018ul}%
  \BibitemOpen
  \bibfield  {author} {\bibinfo {author} {\bibfnamefont {M.}~\bibnamefont
  {Alizadeh}}, \bibinfo {author} {\bibfnamefont {J.}~\bibnamefont
  {Fernández-Marqués}}, \bibinfo {author} {\bibfnamefont {N.~D.}\
  \bibnamefont {Lane}}, \ and\ \bibinfo {author} {\bibfnamefont
  {Y.}~\bibnamefont {Gal}},\ }in\ \href
  {https://openreview.net/forum?id=rJfUCoR5KX} {\emph {\bibinfo {booktitle}
  {International Conference on Learning Representations}}}\ (\bibinfo {year}
  {2019})\BibitemShut {NoStop}%
\bibitem [{\citenamefont {Deutsch}\ and\ \citenamefont
  {Jozsa}(1992)}]{deutsch1992rapid}%
  \BibitemOpen
  \bibfield  {author} {\bibinfo {author} {\bibfnamefont {D.}~\bibnamefont
  {Deutsch}}\ and\ \bibinfo {author} {\bibfnamefont {R.}~\bibnamefont
  {Jozsa}},\ }\href@noop {} {\bibfield  {journal} {\bibinfo  {journal} {Proc.
  R. Soc. Lond. A}\ }\textbf {\bibinfo {volume} {439}},\ \bibinfo {pages} {553}
  (\bibinfo {year} {1992})}\BibitemShut {NoStop}%
\bibitem [{\citenamefont {Grover}(1996)}]{grover1996fast}%
  \BibitemOpen
  \bibfield  {author} {\bibinfo {author} {\bibfnamefont {L.~K.}\ \bibnamefont
  {Grover}},\ }in\ \href@noop {} {\emph {\bibinfo {booktitle} {Proceedings of
  the twenty-eighth annual ACM symposium on Theory of computing}}}\ (\bibinfo
  {organization} {ACM},\ \bibinfo {year} {1996})\ pp.\ \bibinfo {pages}
  {212--219}\BibitemShut {NoStop}%
\bibitem [{\citenamefont {Shor}(1999)}]{shor1999polynomial}%
  \BibitemOpen
  \bibfield  {author} {\bibinfo {author} {\bibfnamefont {P.~W.}\ \bibnamefont
  {Shor}},\ }\href@noop {} {\bibfield  {journal} {\bibinfo  {journal} {SIAM
  review}\ }\textbf {\bibinfo {volume} {41}},\ \bibinfo {pages} {303} (\bibinfo
  {year} {1999})}\BibitemShut {NoStop}%
\bibitem [{\citenamefont {Bravyi}\ \emph {et~al.}(2017)\citenamefont {Bravyi},
  \citenamefont {Gosset},\ and\ \citenamefont {Koenig}}]{bravyi2017quantum}%
  \BibitemOpen
  \bibfield  {author} {\bibinfo {author} {\bibfnamefont {S.}~\bibnamefont
  {Bravyi}}, \bibinfo {author} {\bibfnamefont {D.}~\bibnamefont {Gosset}}, \
  and\ \bibinfo {author} {\bibfnamefont {R.}~\bibnamefont {Koenig}},\
  }\href@noop {} {\bibfield  {journal} {\bibinfo  {journal} {arXiv preprint
  arXiv:1704.00690}\ } (\bibinfo {year} {2017})}\BibitemShut {NoStop}%
\bibitem [{\citenamefont {Schuld}\ \emph {et~al.}(2014)\citenamefont {Schuld},
  \citenamefont {Sinayskiy},\ and\ \citenamefont
  {Petruccione}}]{schuld2014quest}%
  \BibitemOpen
  \bibfield  {author} {\bibinfo {author} {\bibfnamefont {M.}~\bibnamefont
  {Schuld}}, \bibinfo {author} {\bibfnamefont {I.}~\bibnamefont {Sinayskiy}}, \
  and\ \bibinfo {author} {\bibfnamefont {F.}~\bibnamefont {Petruccione}},\
  }\href@noop {} {\bibfield  {journal} {\bibinfo  {journal} {Quantum
  Information Processing}\ }\textbf {\bibinfo {volume} {13}},\ \bibinfo {pages}
  {2567} (\bibinfo {year} {2014})}\BibitemShut {NoStop}%
\bibitem [{\citenamefont {Wan}\ \emph {et~al.}(2018)\citenamefont {Wan},
  \citenamefont {Liu}, \citenamefont {Dahlsten},\ and\ \citenamefont
  {Kim}}]{wan2018learning}%
  \BibitemOpen
  \bibfield  {author} {\bibinfo {author} {\bibfnamefont {K.~H.}\ \bibnamefont
  {Wan}}, \bibinfo {author} {\bibfnamefont {F.}~\bibnamefont {Liu}}, \bibinfo
  {author} {\bibfnamefont {O.}~\bibnamefont {Dahlsten}}, \ and\ \bibinfo
  {author} {\bibfnamefont {M.}~\bibnamefont {Kim}},\ }\href@noop {} {\bibfield
  {journal} {\bibinfo  {journal} {arXiv preprint arXiv:1806.10448}\ } (\bibinfo
  {year} {2018})}\BibitemShut {NoStop}%
\bibitem [{\citenamefont {Morales}\ \emph {et~al.}(2018)\citenamefont
  {Morales}, \citenamefont {Tlyachev},\ and\ \citenamefont
  {Biamonte}}]{morales2018variationally}%
  \BibitemOpen
  \bibfield  {author} {\bibinfo {author} {\bibfnamefont {M.~E.}\ \bibnamefont
  {Morales}}, \bibinfo {author} {\bibfnamefont {T.}~\bibnamefont {Tlyachev}}, \
  and\ \bibinfo {author} {\bibfnamefont {J.}~\bibnamefont {Biamonte}},\
  }\href@noop {} {\bibfield  {journal} {\bibinfo  {journal} {arXiv preprint
  arXiv:1805.09337}\ } (\bibinfo {year} {2018})}\BibitemShut {NoStop}%
\bibitem [{\citenamefont {Wan}\ \emph {et~al.}(2017)\citenamefont {Wan},
  \citenamefont {Dahlsten}, \citenamefont {Kristj{\'a}nsson}, \citenamefont
  {Gardner},\ and\ \citenamefont {Kim}}]{wan2017quantum}%
  \BibitemOpen
  \bibfield  {author} {\bibinfo {author} {\bibfnamefont {K.~H.}\ \bibnamefont
  {Wan}}, \bibinfo {author} {\bibfnamefont {O.}~\bibnamefont {Dahlsten}},
  \bibinfo {author} {\bibfnamefont {H.}~\bibnamefont {Kristj{\'a}nsson}},
  \bibinfo {author} {\bibfnamefont {R.}~\bibnamefont {Gardner}}, \ and\
  \bibinfo {author} {\bibfnamefont {M.}~\bibnamefont {Kim}},\ }\href@noop {}
  {\bibfield  {journal} {\bibinfo  {journal} {npj Quantum Information}\
  }\textbf {\bibinfo {volume} {3}},\ \bibinfo {pages} {36} (\bibinfo {year}
  {2017})}\BibitemShut {NoStop}%
\bibitem [{\citenamefont {Beer}\ \emph {et~al.}(2019)\citenamefont {Beer},
  \citenamefont {Bondarenko}, \citenamefont {Farrelly}, \citenamefont
  {Osborne}, \citenamefont {Salzmann},\ and\ \citenamefont
  {Wolf}}]{beer2019efficient}%
  \BibitemOpen
  \bibfield  {author} {\bibinfo {author} {\bibfnamefont {K.}~\bibnamefont
  {Beer}}, \bibinfo {author} {\bibfnamefont {D.}~\bibnamefont {Bondarenko}},
  \bibinfo {author} {\bibfnamefont {T.}~\bibnamefont {Farrelly}}, \bibinfo
  {author} {\bibfnamefont {T.~J.}\ \bibnamefont {Osborne}}, \bibinfo {author}
  {\bibfnamefont {R.}~\bibnamefont {Salzmann}}, \ and\ \bibinfo {author}
  {\bibfnamefont {R.}~\bibnamefont {Wolf}},\ }\href@noop {} {\bibfield
  {journal} {\bibinfo  {journal} {arXiv preprint arXiv:1902.10445}\ } (\bibinfo
  {year} {2019})}\BibitemShut {NoStop}%
\bibitem [{\citenamefont {Bergholm}\ \emph {et~al.}(2018)\citenamefont
  {Bergholm}, \citenamefont {Izaac}, \citenamefont {Schuld}, \citenamefont
  {Gogolin},\ and\ \citenamefont {Killoran}}]{bergholm2018pennylane}%
  \BibitemOpen
  \bibfield  {author} {\bibinfo {author} {\bibfnamefont {V.}~\bibnamefont
  {Bergholm}}, \bibinfo {author} {\bibfnamefont {J.}~\bibnamefont {Izaac}},
  \bibinfo {author} {\bibfnamefont {M.}~\bibnamefont {Schuld}}, \bibinfo
  {author} {\bibfnamefont {C.}~\bibnamefont {Gogolin}}, \ and\ \bibinfo
  {author} {\bibfnamefont {N.}~\bibnamefont {Killoran}},\ }\href@noop {}
  {\bibfield  {journal} {\bibinfo  {journal} {arXiv preprint arXiv:1811.04968}\
  } (\bibinfo {year} {2018})}\BibitemShut {NoStop}%
\bibitem [{\citenamefont {Daskin}(2018)}]{daskin2018simple}%
  \BibitemOpen
  \bibfield  {author} {\bibinfo {author} {\bibfnamefont {A.}~\bibnamefont
  {Daskin}},\ }in\ \href@noop {} {\emph {\bibinfo {booktitle} {2018 IEEE
  International Conference on Systems, Man, and Cybernetics (SMC)}}}\ (\bibinfo
  {organization} {IEEE},\ \bibinfo {year} {2018})\ pp.\ \bibinfo {pages}
  {2887--2891}\BibitemShut {NoStop}%
\bibitem [{\citenamefont {Zhao}\ \emph {et~al.}(2018)\citenamefont {Zhao},
  \citenamefont {Sun}, \citenamefont {Feng}, \citenamefont {Zhao},
  \citenamefont {Sui},\ and\ \citenamefont {Xu}}]{zhao2018forecast}%
  \BibitemOpen
  \bibfield  {author} {\bibinfo {author} {\bibfnamefont {J.}~\bibnamefont
  {Zhao}}, \bibinfo {author} {\bibfnamefont {Y.}~\bibnamefont {Sun}}, \bibinfo
  {author} {\bibfnamefont {F.}~\bibnamefont {Feng}}, \bibinfo {author}
  {\bibfnamefont {F.}~\bibnamefont {Zhao}}, \bibinfo {author} {\bibfnamefont
  {D.}~\bibnamefont {Sui}}, \ and\ \bibinfo {author} {\bibfnamefont
  {J.}~\bibnamefont {Xu}},\ }in\ \href@noop {} {\emph {\bibinfo {booktitle}
  {IOP Conference Series: Earth and Environmental Science}}},\ Vol.\ \bibinfo
  {volume} {108}\ (\bibinfo {organization} {IOP Publishing},\ \bibinfo {year}
  {2018})\ p.\ \bibinfo {pages} {032008}\BibitemShut {NoStop}%
\bibitem [{\citenamefont {Farhi}\ \emph {et~al.}()\citenamefont {Farhi},
  \citenamefont {1802.06002},\ and\ \citenamefont {{2018}}}]{Farhi:wv}%
  \BibitemOpen
  \bibfield  {author} {\bibinfo {author} {\bibfnamefont {E.}~\bibnamefont
  {Farhi}}, \bibinfo {author} {\bibfnamefont {H.~N. a. p.~a.}\ \bibnamefont
  {1802.06002}}, \ and\ \bibinfo {author} {\bibnamefont {{2018}}},\ }\href@noop
  {} {\bibinfo  {journal} {arxiv.org}\ }\BibitemShut {NoStop}%
\bibitem [{\citenamefont {McClean}\ \emph {et~al.}(2018)\citenamefont
  {McClean}, \citenamefont {Boixo}, \citenamefont {Smelyanskiy}, \citenamefont
  {Babbush},\ and\ \citenamefont {Neven}}]{McClean:2018um}%
  \BibitemOpen
\bibfield  {journal} {  }\bibfield  {author} {\bibinfo {author} {\bibfnamefont
  {J.~R.}\ \bibnamefont {McClean}}, \bibinfo {author} {\bibfnamefont
  {S.}~\bibnamefont {Boixo}}, \bibinfo {author} {\bibfnamefont {V.~N.}\
  \bibnamefont {Smelyanskiy}}, \bibinfo {author} {\bibfnamefont
  {R.}~\bibnamefont {Babbush}}, \ and\ \bibinfo {author} {\bibfnamefont
  {H.}~\bibnamefont {Neven}},\ }\href@noop {} {\bibfield  {journal} {\bibinfo
  {journal} {Nature Communications}\ }\textbf {\bibinfo {volume} {9}},\
  \bibinfo {pages} {4812} (\bibinfo {year} {2018})}\BibitemShut {NoStop}%
\bibitem [{\citenamefont {Verdon}\ \emph {et~al.}(2018)\citenamefont {Verdon},
  \citenamefont {Pye},\ and\ \citenamefont {Broughton}}]{verdon2018universal}%
  \BibitemOpen
  \bibfield  {author} {\bibinfo {author} {\bibfnamefont {G.}~\bibnamefont
  {Verdon}}, \bibinfo {author} {\bibfnamefont {J.}~\bibnamefont {Pye}}, \ and\
  \bibinfo {author} {\bibfnamefont {M.}~\bibnamefont {Broughton}},\ }\href@noop
  {} {\bibfield  {journal} {\bibinfo  {journal} {arXiv preprint
  arXiv:1806.09729}\ } (\bibinfo {year} {2018})}\BibitemShut {NoStop}%
\bibitem [{\citenamefont {Gilyen}\ \emph {et~al.}(2017)\citenamefont {Gilyen},
  \citenamefont {Arunachalam},\ and\ \citenamefont {Wiebe}}]{GilyenASW17}%
  \BibitemOpen
  \bibfield  {author} {\bibinfo {author} {\bibfnamefont {A.}~\bibnamefont
  {Gilyen}}, \bibinfo {author} {\bibfnamefont {S.}~\bibnamefont {Arunachalam}},
  \ and\ \bibinfo {author} {\bibfnamefont {N.}~\bibnamefont {Wiebe}},\
  }\href@noop {} {\bibfield  {journal} {\bibinfo  {journal} {arXiv preprint
  arXiv:1711.00465}\ } (\bibinfo {year} {2017})}\BibitemShut {NoStop}%
\bibitem [{\citenamefont {{Jordan}}(2005)}]{Jordan05}%
  \BibitemOpen
  \bibfield  {author} {\bibinfo {author} {\bibfnamefont {S.~P.}\ \bibnamefont
  {{Jordan}}},\ }\href {\doibase 10.1103/PhysRevLett.95.050501} {\bibfield
  {journal} {\bibinfo  {journal} {\prl}\ }\textbf {\bibinfo {volume} {95}},\
  \bibinfo {pages} {050501} (\bibinfo {year} {2005})}\BibitemShut {NoStop}%
\bibitem [{\citenamefont {Ricks}\ and\ \citenamefont
  {Ventura}(2004)}]{ricks2004training}%
  \BibitemOpen
  \bibfield  {author} {\bibinfo {author} {\bibfnamefont {B.}~\bibnamefont
  {Ricks}}\ and\ \bibinfo {author} {\bibfnamefont {D.}~\bibnamefont
  {Ventura}},\ }in\ \href@noop {} {\emph {\bibinfo {booktitle} {Advances in
  neural information processing systems}}}\ (\bibinfo {year} {2004})\ pp.\
  \bibinfo {pages} {1019--1026}\BibitemShut {NoStop}%
\bibitem [{\citenamefont {Kotsiantis}\ \emph {et~al.}(2006)\citenamefont
  {Kotsiantis}, \citenamefont {Zaharakis},\ and\ \citenamefont
  {Pintelas}}]{kotsiantis2006machine}%
  \BibitemOpen
  \bibfield  {author} {\bibinfo {author} {\bibfnamefont {S.~B.}\ \bibnamefont
  {Kotsiantis}}, \bibinfo {author} {\bibfnamefont {I.~D.}\ \bibnamefont
  {Zaharakis}}, \ and\ \bibinfo {author} {\bibfnamefont {P.~E.}\ \bibnamefont
  {Pintelas}},\ }\href@noop {} {\bibfield  {journal} {\bibinfo  {journal}
  {Artificial Intelligence Review}\ }\textbf {\bibinfo {volume} {26}},\
  \bibinfo {pages} {159} (\bibinfo {year} {2006})}\BibitemShut {NoStop}%
\bibitem [{\citenamefont {Courbariaux}\ \emph {et~al.}(2015)\citenamefont
  {Courbariaux}, \citenamefont {Bengio},\ and\ \citenamefont
  {David}}]{courbariaux2015binaryconnect}%
  \BibitemOpen
  \bibfield  {author} {\bibinfo {author} {\bibfnamefont {M.}~\bibnamefont
  {Courbariaux}}, \bibinfo {author} {\bibfnamefont {Y.}~\bibnamefont {Bengio}},
  \ and\ \bibinfo {author} {\bibfnamefont {J.-P.}\ \bibnamefont {David}},\ }in\
  \href@noop {} {\emph {\bibinfo {booktitle} {Advances in neural information
  processing systems}}}\ (\bibinfo {year} {2015})\ pp.\ \bibinfo {pages}
  {3123--3131}\BibitemShut {NoStop}%
\bibitem [{\citenamefont {Livni}\ \emph {et~al.}(2014)\citenamefont {Livni},
  \citenamefont {Shalev-Shwartz},\ and\ \citenamefont
  {Shamir}}]{livni2014computational}%
  \BibitemOpen
  \bibfield  {author} {\bibinfo {author} {\bibfnamefont {R.}~\bibnamefont
  {Livni}}, \bibinfo {author} {\bibfnamefont {S.}~\bibnamefont
  {Shalev-Shwartz}}, \ and\ \bibinfo {author} {\bibfnamefont {O.}~\bibnamefont
  {Shamir}},\ }in\ \href@noop {} {\emph {\bibinfo {booktitle} {Advances in
  neural information processing systems}}}\ (\bibinfo {year} {2014})\ pp.\
  \bibinfo {pages} {855--863}\BibitemShut {NoStop}%
\bibitem [{\citenamefont {Nielsen}\ and\ \citenamefont
  {Chuang}(2002)}]{nielsen2002quantum}%
  \BibitemOpen
  \bibfield  {author} {\bibinfo {author} {\bibfnamefont {M.~A.}\ \bibnamefont
  {Nielsen}}\ and\ \bibinfo {author} {\bibfnamefont {I.}~\bibnamefont
  {Chuang}},\ }\href@noop {} {\emph {\bibinfo {title} {Quantum computation and
  quantum information}}}\ (\bibinfo  {publisher} {AAPT},\ \bibinfo {year}
  {2002})\BibitemShut {NoStop}%
\bibitem [{\citenamefont {Barenco}\ \emph {et~al.}(1995)\citenamefont
  {Barenco}, \citenamefont {Bennett}, \citenamefont {Cleve}, \citenamefont
  {DiVincenzo}, \citenamefont {Margolus}, \citenamefont {Shor}, \citenamefont
  {Sleator}, \citenamefont {Smolin},\ and\ \citenamefont
  {Weinfurter}}]{barenco1995elementary}%
  \BibitemOpen
  \bibfield  {author} {\bibinfo {author} {\bibfnamefont {A.}~\bibnamefont
  {Barenco}}, \bibinfo {author} {\bibfnamefont {C.~H.}\ \bibnamefont
  {Bennett}}, \bibinfo {author} {\bibfnamefont {R.}~\bibnamefont {Cleve}},
  \bibinfo {author} {\bibfnamefont {D.~P.}\ \bibnamefont {DiVincenzo}},
  \bibinfo {author} {\bibfnamefont {N.}~\bibnamefont {Margolus}}, \bibinfo
  {author} {\bibfnamefont {P.}~\bibnamefont {Shor}}, \bibinfo {author}
  {\bibfnamefont {T.}~\bibnamefont {Sleator}}, \bibinfo {author} {\bibfnamefont
  {J.~A.}\ \bibnamefont {Smolin}}, \ and\ \bibinfo {author} {\bibfnamefont
  {H.}~\bibnamefont {Weinfurter}},\ }\href@noop {} {\bibfield  {journal}
  {\bibinfo  {journal} {Physical Review A}\ }\textbf {\bibinfo {volume} {52}},\
  \bibinfo {pages} {3457} (\bibinfo {year} {1995})}\BibitemShut {NoStop}%
\bibitem [{\citenamefont {Wootters}\ and\ \citenamefont
  {Zurek}(1982)}]{wootters1982single}%
  \BibitemOpen
  \bibfield  {author} {\bibinfo {author} {\bibfnamefont {W.~K.}\ \bibnamefont
  {Wootters}}\ and\ \bibinfo {author} {\bibfnamefont {W.~H.}\ \bibnamefont
  {Zurek}},\ }\href@noop {} {\bibfield  {journal} {\bibinfo  {journal}
  {Nature}\ }\textbf {\bibinfo {volume} {299}},\ \bibinfo {pages} {802}
  (\bibinfo {year} {1982})}\BibitemShut {NoStop}%
\bibitem [{\citenamefont {Grover}(1998)}]{grover1998quantum}%
  \BibitemOpen
  \bibfield  {author} {\bibinfo {author} {\bibfnamefont {L.~K.}\ \bibnamefont
  {Grover}},\ }\href@noop {} {\bibfield  {journal} {\bibinfo  {journal}
  {Physical Review Letters}\ }\textbf {\bibinfo {volume} {80}},\ \bibinfo
  {pages} {4329} (\bibinfo {year} {1998})}\BibitemShut {NoStop}%
\bibitem [{\citenamefont {Boyer}\ \emph {et~al.}(1998)\citenamefont {Boyer},
  \citenamefont {Brassard}, \citenamefont {H{\o}yer},\ and\ \citenamefont
  {Tapp}}]{boyer1998tight}%
  \BibitemOpen
  \bibfield  {author} {\bibinfo {author} {\bibfnamefont {M.}~\bibnamefont
  {Boyer}}, \bibinfo {author} {\bibfnamefont {G.}~\bibnamefont {Brassard}},
  \bibinfo {author} {\bibfnamefont {P.}~\bibnamefont {H{\o}yer}}, \ and\
  \bibinfo {author} {\bibfnamefont {A.}~\bibnamefont {Tapp}},\ }\href@noop {}
  {\bibfield  {journal} {\bibinfo  {journal} {Fortschritte der Physik: Progress
  of Physics}\ }\textbf {\bibinfo {volume} {46}},\ \bibinfo {pages} {493}
  (\bibinfo {year} {1998})}\BibitemShut {NoStop}%
\bibitem [{\citenamefont {Huang}(2003)}]{huang2003learning}%
  \BibitemOpen
  \bibfield  {author} {\bibinfo {author} {\bibfnamefont {G.-B.}\ \bibnamefont
  {Huang}},\ }\href@noop {} {\bibfield  {journal} {\bibinfo  {journal} {IEEE
  Transactions on Neural Networks}\ }\textbf {\bibinfo {volume} {14}},\
  \bibinfo {pages} {274} (\bibinfo {year} {2003})}\BibitemShut {NoStop}%
\bibitem [{\citenamefont {Mocanu}\ \emph {et~al.}(2018)\citenamefont {Mocanu},
  \citenamefont {Mocanu}, \citenamefont {Stone}, \citenamefont {Nguyen},
  \citenamefont {Gibescu},\ and\ \citenamefont {Liotta}}]{mocanu2018scalable}%
  \BibitemOpen
  \bibfield  {author} {\bibinfo {author} {\bibfnamefont {D.~C.}\ \bibnamefont
  {Mocanu}}, \bibinfo {author} {\bibfnamefont {E.}~\bibnamefont {Mocanu}},
  \bibinfo {author} {\bibfnamefont {P.}~\bibnamefont {Stone}}, \bibinfo
  {author} {\bibfnamefont {P.~H.}\ \bibnamefont {Nguyen}}, \bibinfo {author}
  {\bibfnamefont {M.}~\bibnamefont {Gibescu}}, \ and\ \bibinfo {author}
  {\bibfnamefont {A.}~\bibnamefont {Liotta}},\ }\href@noop {} {\bibfield
  {journal} {\bibinfo  {journal} {Nature Communications}\ }\textbf {\bibinfo
  {volume} {9}},\ \bibinfo {pages} {2383} (\bibinfo {year} {2018})}\BibitemShut
  {NoStop}%
\bibitem [{\citenamefont {Glorot}\ \emph {et~al.}(2011)\citenamefont {Glorot},
  \citenamefont {Bordes},\ and\ \citenamefont {Bengio}}]{glorot2011deep}%
  \BibitemOpen
  \bibfield  {author} {\bibinfo {author} {\bibfnamefont {X.}~\bibnamefont
  {Glorot}}, \bibinfo {author} {\bibfnamefont {A.}~\bibnamefont {Bordes}}, \
  and\ \bibinfo {author} {\bibfnamefont {Y.}~\bibnamefont {Bengio}},\ }in\
  \href@noop {} {\emph {\bibinfo {booktitle} {Proceedings of the fourteenth
  international conference on artificial intelligence and statistics}}}\
  (\bibinfo {year} {2011})\ pp.\ \bibinfo {pages} {315--323}\BibitemShut
  {NoStop}%
\bibitem [{\citenamefont {{Bengio}}\ \emph {et~al.}(1994)\citenamefont
  {{Bengio}}, \citenamefont {{Simard}},\ and\ \citenamefont
  {{Frasconi}}}]{bengio1994learning}%
  \BibitemOpen
  \bibfield  {author} {\bibinfo {author} {\bibfnamefont {Y.}~\bibnamefont
  {{Bengio}}}, \bibinfo {author} {\bibfnamefont {P.}~\bibnamefont {{Simard}}},
  \ and\ \bibinfo {author} {\bibfnamefont {P.}~\bibnamefont {{Frasconi}}},\
  }\href {\doibase 10.1109/72.279181} {\bibfield  {journal} {\bibinfo
  {journal} {IEEE Transactions on Neural Networks}\ }\textbf {\bibinfo {volume}
  {5}},\ \bibinfo {pages} {157} (\bibinfo {year} {1994})}\BibitemShut {NoStop}%
\bibitem [{\citenamefont {{Yi Shang}}\ and\ \citenamefont
  {{Wah}}(1996)}]{Shang485892}%
  \BibitemOpen
  \bibfield  {author} {\bibinfo {author} {\bibnamefont {{Yi Shang}}}\ and\
  \bibinfo {author} {\bibfnamefont {B.~W.}\ \bibnamefont {{Wah}}},\ }\href
  {\doibase 10.1109/2.485892} {\bibfield  {journal} {\bibinfo  {journal}
  {Computer}\ }\textbf {\bibinfo {volume} {29}},\ \bibinfo {pages} {45}
  (\bibinfo {year} {1996})}\BibitemShut {NoStop}%
\bibitem [{\citenamefont {Watrous}\ and\ \citenamefont
  {Kuhn}(1992)}]{watrous1992induction}%
  \BibitemOpen
  \bibfield  {author} {\bibinfo {author} {\bibfnamefont {R.~L.}\ \bibnamefont
  {Watrous}}\ and\ \bibinfo {author} {\bibfnamefont {G.~M.}\ \bibnamefont
  {Kuhn}},\ }\href@noop {} {\bibfield  {journal} {\bibinfo  {journal} {Neural
  Computation}\ }\textbf {\bibinfo {volume} {4}},\ \bibinfo {pages} {406}
  (\bibinfo {year} {1992})}\BibitemShut {NoStop}%
\bibitem [{\citenamefont {{Graves}}\ \emph {et~al.}(2009)\citenamefont
  {{Graves}}, \citenamefont {{Liwicki}}, \citenamefont {{Fernández}},
  \citenamefont {{Bertolami}}, \citenamefont {{Bunke}},\ and\ \citenamefont
  {{Schmidhuber}}}]{graves2008novel}%
  \BibitemOpen
  \bibfield  {author} {\bibinfo {author} {\bibfnamefont {A.}~\bibnamefont
  {{Graves}}}, \bibinfo {author} {\bibfnamefont {M.}~\bibnamefont {{Liwicki}}},
  \bibinfo {author} {\bibfnamefont {S.}~\bibnamefont {{Fernández}}}, \bibinfo
  {author} {\bibfnamefont {R.}~\bibnamefont {{Bertolami}}}, \bibinfo {author}
  {\bibfnamefont {H.}~\bibnamefont {{Bunke}}}, \ and\ \bibinfo {author}
  {\bibfnamefont {J.}~\bibnamefont {{Schmidhuber}}},\ }\href {\doibase
  10.1109/TPAMI.2008.137} {\bibfield  {journal} {\bibinfo  {journal} {IEEE
  Transactions on Pattern Analysis and Machine Intelligence}\ }\textbf
  {\bibinfo {volume} {31}},\ \bibinfo {pages} {855} (\bibinfo {year}
  {2009})}\BibitemShut {NoStop}%
\bibitem [{\citenamefont {Sutskever}\ \emph {et~al.}(2011)\citenamefont
  {Sutskever}, \citenamefont {Martens},\ and\ \citenamefont
  {Hinton}}]{sutskever2011generating}%
  \BibitemOpen
  \bibfield  {author} {\bibinfo {author} {\bibfnamefont {I.}~\bibnamefont
  {Sutskever}}, \bibinfo {author} {\bibfnamefont {J.}~\bibnamefont {Martens}},
  \ and\ \bibinfo {author} {\bibfnamefont {G.~E.}\ \bibnamefont {Hinton}},\
  }in\ \href@noop {} {\emph {\bibinfo {booktitle} {Proceedings of the 28th
  International Conference on Machine Learning (ICML-11)}}}\ (\bibinfo {year}
  {2011})\ pp.\ \bibinfo {pages} {1017--1024}\BibitemShut {NoStop}%
\bibitem [{\citenamefont {Hochreiter}(1998)}]{hochreiter1998vanishing}%
  \BibitemOpen
  \bibfield  {author} {\bibinfo {author} {\bibfnamefont {S.}~\bibnamefont
  {Hochreiter}},\ }\href@noop {} {\bibfield  {journal} {\bibinfo  {journal}
  {International Journal of Uncertainty, Fuzziness and Knowledge-Based
  Systems}\ }\textbf {\bibinfo {volume} {6}},\ \bibinfo {pages} {107} (\bibinfo
  {year} {1998})}\BibitemShut {NoStop}%
\bibitem [{\citenamefont {Pascanu}\ \emph {et~al.}(2013)\citenamefont
  {Pascanu}, \citenamefont {Mikolov},\ and\ \citenamefont
  {Bengio}}]{pascanu2013difficulty}%
  \BibitemOpen
  \bibfield  {author} {\bibinfo {author} {\bibfnamefont {R.}~\bibnamefont
  {Pascanu}}, \bibinfo {author} {\bibfnamefont {T.}~\bibnamefont {Mikolov}}, \
  and\ \bibinfo {author} {\bibfnamefont {Y.}~\bibnamefont {Bengio}},\ }in\
  \href@noop {} {\emph {\bibinfo {booktitle} {International conference on
  machine learning}}}\ (\bibinfo {year} {2013})\ pp.\ \bibinfo {pages}
  {1310--1318}\BibitemShut {NoStop}%
\bibitem [{\citenamefont {Srivastava}\ \emph {et~al.}(2014)\citenamefont
  {Srivastava}, \citenamefont {Hinton}, \citenamefont {Krizhevsky},
  \citenamefont {Sutskever},\ and\ \citenamefont
  {Salakhutdinov}}]{srivastava2014dropout}%
  \BibitemOpen
  \bibfield  {author} {\bibinfo {author} {\bibfnamefont {N.}~\bibnamefont
  {Srivastava}}, \bibinfo {author} {\bibfnamefont {G.}~\bibnamefont {Hinton}},
  \bibinfo {author} {\bibfnamefont {A.}~\bibnamefont {Krizhevsky}}, \bibinfo
  {author} {\bibfnamefont {I.}~\bibnamefont {Sutskever}}, \ and\ \bibinfo
  {author} {\bibfnamefont {R.}~\bibnamefont {Salakhutdinov}},\ }\href@noop {}
  {\bibfield  {journal} {\bibinfo  {journal} {The Journal of Machine Learning
  Research}\ }\textbf {\bibinfo {volume} {15}},\ \bibinfo {pages} {1929}
  (\bibinfo {year} {2014})}\BibitemShut {NoStop}%
\bibitem [{\citenamefont {Shrestha}\ and\ \citenamefont
  {Mahmood}(2019)}]{shrestha2019review}%
  \BibitemOpen
  \bibfield  {author} {\bibinfo {author} {\bibfnamefont {A.}~\bibnamefont
  {Shrestha}}\ and\ \bibinfo {author} {\bibfnamefont {A.}~\bibnamefont
  {Mahmood}},\ }\href@noop {} {\bibfield  {journal} {\bibinfo  {journal} {IEEE
  Access}\ }\textbf {\bibinfo {volume} {7}},\ \bibinfo {pages} {53040}
  (\bibinfo {year} {2019})}\BibitemShut {NoStop}%
\end{thebibliography}%

\appendix
\onecolumngrid

\newpage
 
\section{Circuit implementation of the full phase estumation training cycle \label{app:fulltrain}}
In the main text, Section \ref{fulltrain} presented a schematic overview of the complete cycle of the quantum training via phase estimation for a single QBN. Here, we provide the circuit implementation for the the full training, which consists of $k^*$ training cycles (c.f. Figure \ref{fig:cycle}).

\begin{figure}[h]
\includegraphics[width=0.95\linewidth]{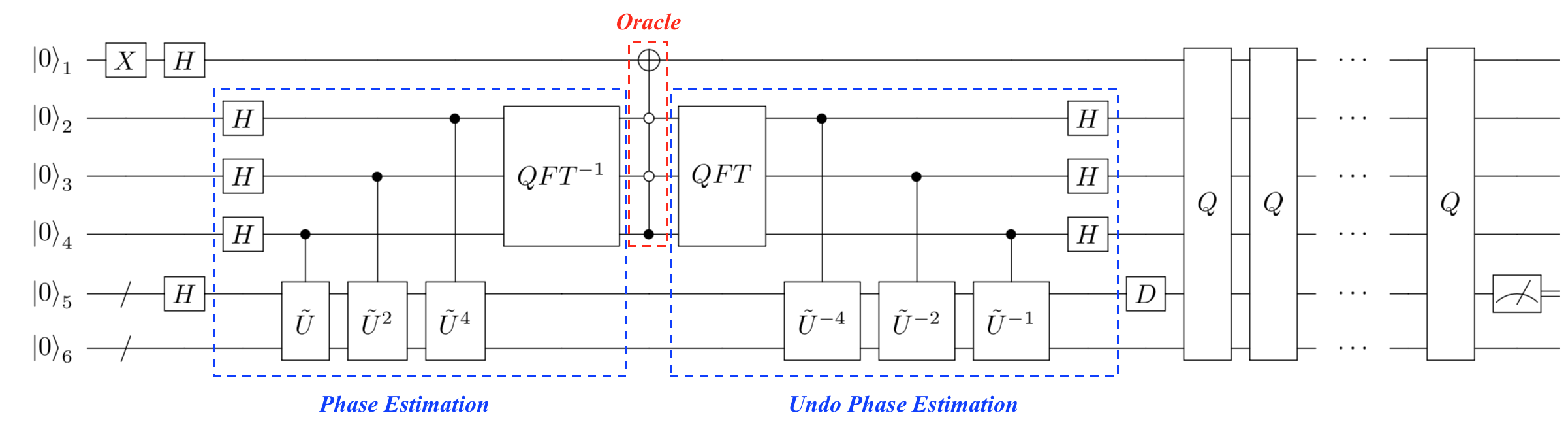}
\caption{\textit{Circuit implementation for the full training.}}
\label{fig:cycle2}
\end{figure}

As depicted in \ref{fig:cycle2}, from top to bottom, we first identify the circuit's input qubits (all initialized in state $|0\>$). The very first qubit $\ket{0}_1$ plays the role of a control for the binary marking oracle $O_{\pm 1}$. As described in the main text, the oracle decides whether the quality of a weight string is above a certain threshold $N_t$, and if, it so changes the sign of the according string. To do so, the control qubit of the oracle is to be transformed into state $|-\>$ to perform the threshold comparison correctly (we explain the functioning of the oracle below). The next three input qubits $\ket{0}_2-\ket{0}_4$ are the ancillary control systems for the phase estimation. In the schematic circuit discussed here this means that the phase estimation subroutine estimates $t=3$ digits in binary of the number $\phi$ in the phase $e^{i \pi 2 \phi}$ (up to some error described in Section \ref{perfPE}). The remaining qubits play the following role: the multi-qubit register $|0\>_5$ represents the weight register, which is transformed into the coherent superposition $|W\>=\frac{1}{\sqrt{\tilde{N}}} \sum |\underline{w}\>$ of all weight strings. The last qubits with index 6 are the remaining inputs (training data and ancillas for saving the output of the neural network).  The training inputs and desired outputs are all initialized in zero. As we will see later based on concrete examples, the unitary action $\tilde{U}$ includes changing the training input qubits into the intended training pairs.\\

The first subroutine in the blue box executed the phase estimation. The PE algorithm first initializes the $t=3$ control qubits in the $|+\>$ state. Each qubit controls $2^{j-1}$ uses of the full marking unitary $\tilde{U}$ (see Section \ref{markingtheweights} for more details) on the inputs of $\tilde{U}$. The circuit implementation of $\tilde{U}$ is further discussed in Appendix \ref{app:qbnnexp}. Then, the inverse quantum Fourier transform (QFT) converts the control ancillas into the state $|\phi_1 \phi_2 \phi_3 \>$, which represents a truncated binary encoding of the factor $N_i / 2n$ into the qubit registers. \\

Next, the first red box implements the oracle $O_{\pm 1}$. The function of the oracle is to add a minus sign to the states $|\phi_1 \phi_2 \phi_3 \>$, which encode the quality $N_i$ of the weight string (in binary). In case the $N_i$ supersedes the threshold $N_t$, the state is changed as $|\phi_1 \phi_2 \phi_3 \> \mapsto -|\phi_1 \phi_2 \phi_3 \>$. In this specific example depicted in Figure \ref{fig:cycle2}, the threshold is set to be $N_t=n$, (namely the optimal weight should be good for all the inputs). When converting to binary digits this condition translates to $\phi_1=1, \phi_2=0, \phi_3=0 $.  Therefore we apply a multi-controlled gate to realise the sign flip: only when $\phi_1=1, \phi_2=0,\phi_3=0 \>$ the controlled-NOT gate gets executed on the ancilla (which has been pre-set to $|-\>$ by applying X and a Hadamard gate on it). The full routine is thereafter to be uncomputed, with the purpose of decoupling the marked weight state $|\hat{W}\>=\frac{1}{\sqrt{\tilde{N}}}\sum (-1)^{N_i\geq N_t} |\underline{w}\>$ from the other systems. The uncomputation of PE is depicted by the second blue box.\\

After the decoupling of the weight state, the marked superposition $|\hat{W}\>$ undergoes amplitude amplification: the diffusion operator $D=-H \Lambda_0 H$, depicted by the second red box, amplifies the amplitude of the marked states and dampens the unmarked elements. The full Grover cycle - marking and amplifying - is then collated into one box, called $Q$, which is repeated as described in \ref{bsearch}. This yields an optimal weight string, up to some errors induced by the precision $t$ of the phase estimation algorithm and $\delta$ of the binary search. 

\section{Implementations of QBNN examples}\label{app:qbnnexp}
Section \ref{app:fulltrain} gave a schematic overview of the full training cycle. We now going into detail how the phase accumulation $\tilde{U}$ is implemented as a quantum circuit. We illustrate this with concrete examples below.

\subsection{Single neuron with 2 weights and 2 inputs \label{app:exp22}}
\begin{figure}[h!]
\includegraphics[width=0.3\linewidth]{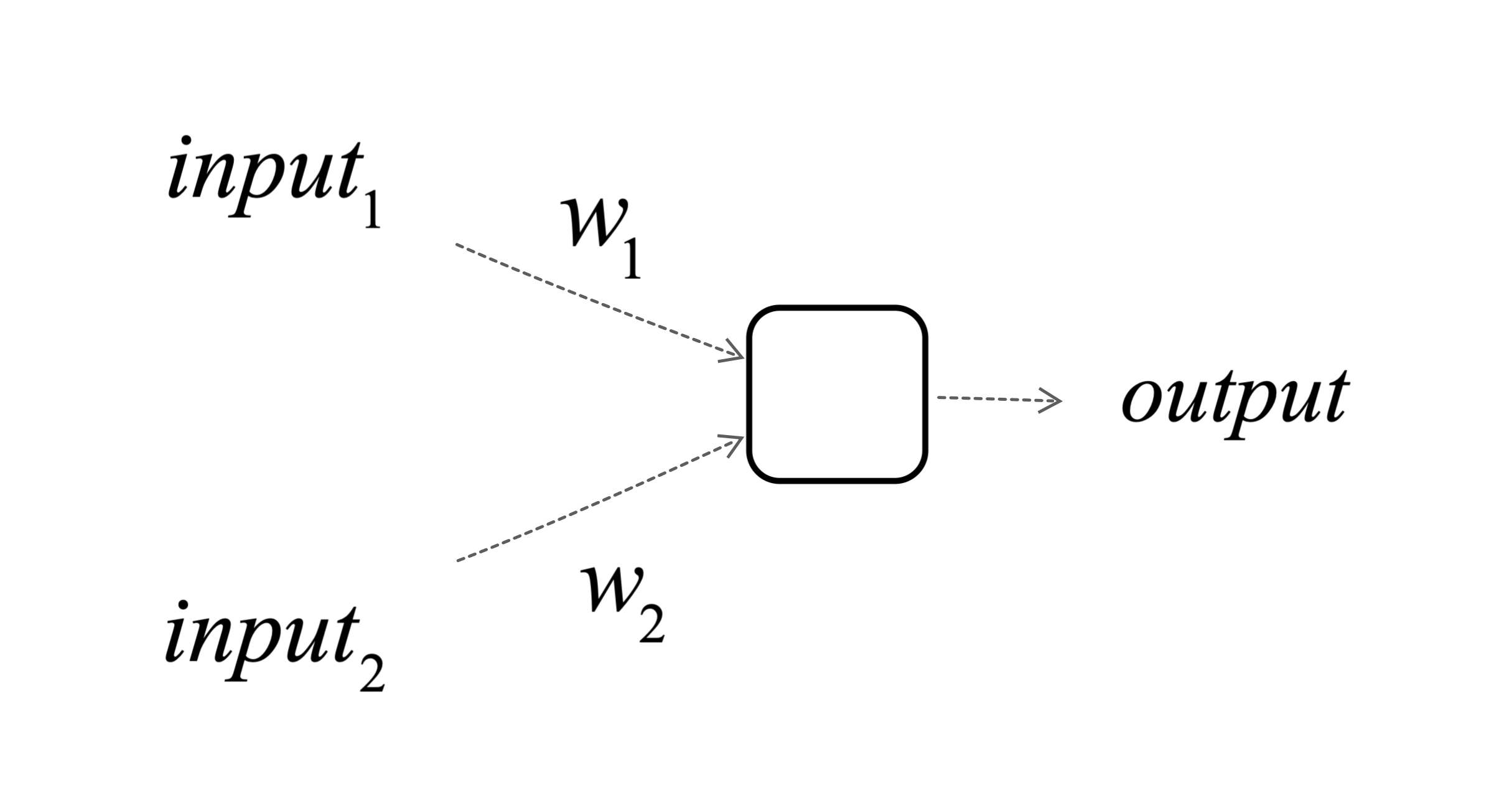}
\caption{Single neuron with 2 inputs and 2 weights.}
\label{fig:QCFON6}
\end{figure}

\begin{figure}[h!]
\includegraphics[width=0.9\linewidth]{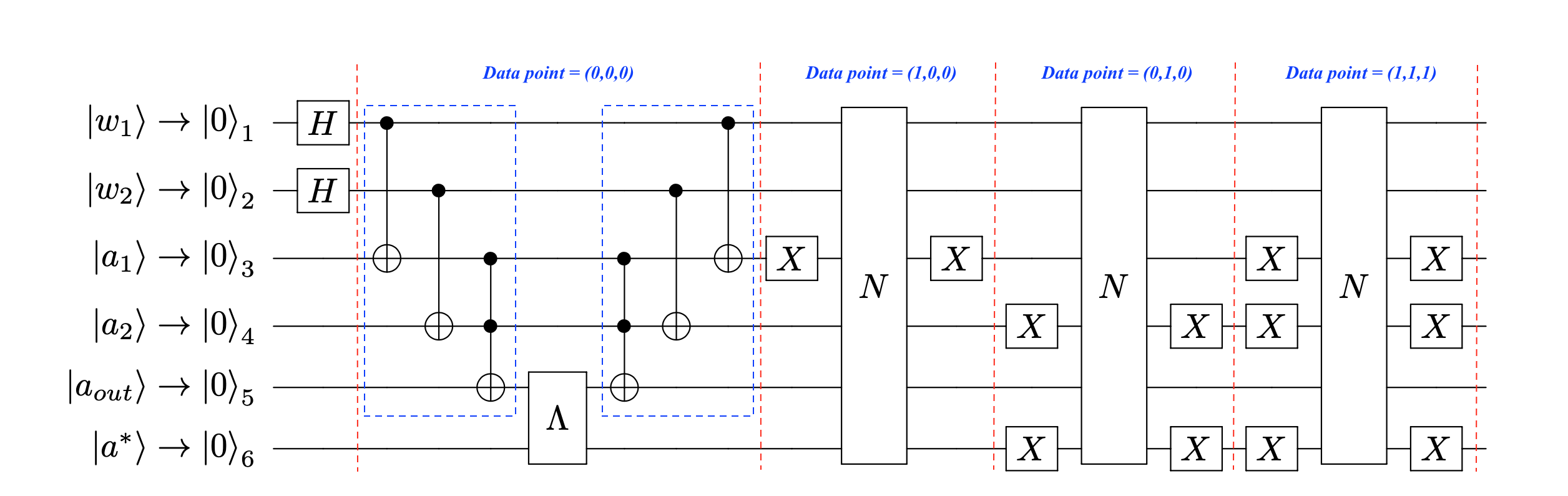}
\caption{Quantum circuit implementation of the phase accumulation sub-routine for the single neuron with 2 inputs and 2 weights.}

\label{fig:QCFON62}
\end{figure}

We begin with studying an elementary instance of a network, namely a single QBN with two inputs and two weights. Figure \ref{fig:QCFON6}
 illustrates the example, together with a circuit implementation of the phase accumulation subroutine--see figure \ref{fig:QCFON62}
.

In this the circuit implementation, the 6 input qubits are (from top to bottom): the two weights $|w_1\>$ and $|w_2\>$, the two training inputs $|a_1\>$ and $|a_2\>$, the ancilla storing the output of the neuron computation, and the desired output $|a^*\>$. The initial training datapoint is $(a_1,a_2,a^*)=(0,0,0)$. In the first step, the Hadamard transformations create the coherent superposition of all possible weight strings, $|W\>=\frac{1}{2} (|00\>+|01\>+|01\>+|11\>)$ (the weight strings are simply $|w\>=|w_1 w_2\>$). Next, the circuit in the first dashed blue box depicts the unitary actions of the neurons: the CNOT gates implements the multiplication between weights and inputs, the following Toffoli gate implements the addition (bitcount operation) of the weighted inputs and the activation function by Sign function. The output is then saved in the ancillary qubit on wire 5. After the oracle $\Lambda$ is called, the uncomputation of the neuron is performed (the circuit in the second blue box). A single phase accumulation cycle is then collectively denoted by the gate $N$.  \\

The full phase accumulation subroutine succeeds by repeating $N$ for $n=4$ times, where $n$ is the number of different training inputs. New input data is initialized in the circuit by applying X gates on $\ket{a_1}$, $\ket{a_2}$, $\ket{a^*}$. In this example, the training set we adopt is: $\{(a_1 , a_2, a^* )\}=\{(0,0,0),(1,0,0),(0,1,1),(1,1,1)\} $ (We exhaust all possible data points as training data, to reasonably make use of the phase accumulation process, for such small example with few inputs. This is also adopted for the other examples in this appendix).

\subsection{Single neuron with 3 weights and 3 inputs \label{singleneur3in}}
\begin{figure}[h!]
\includegraphics[width=0.85\linewidth]{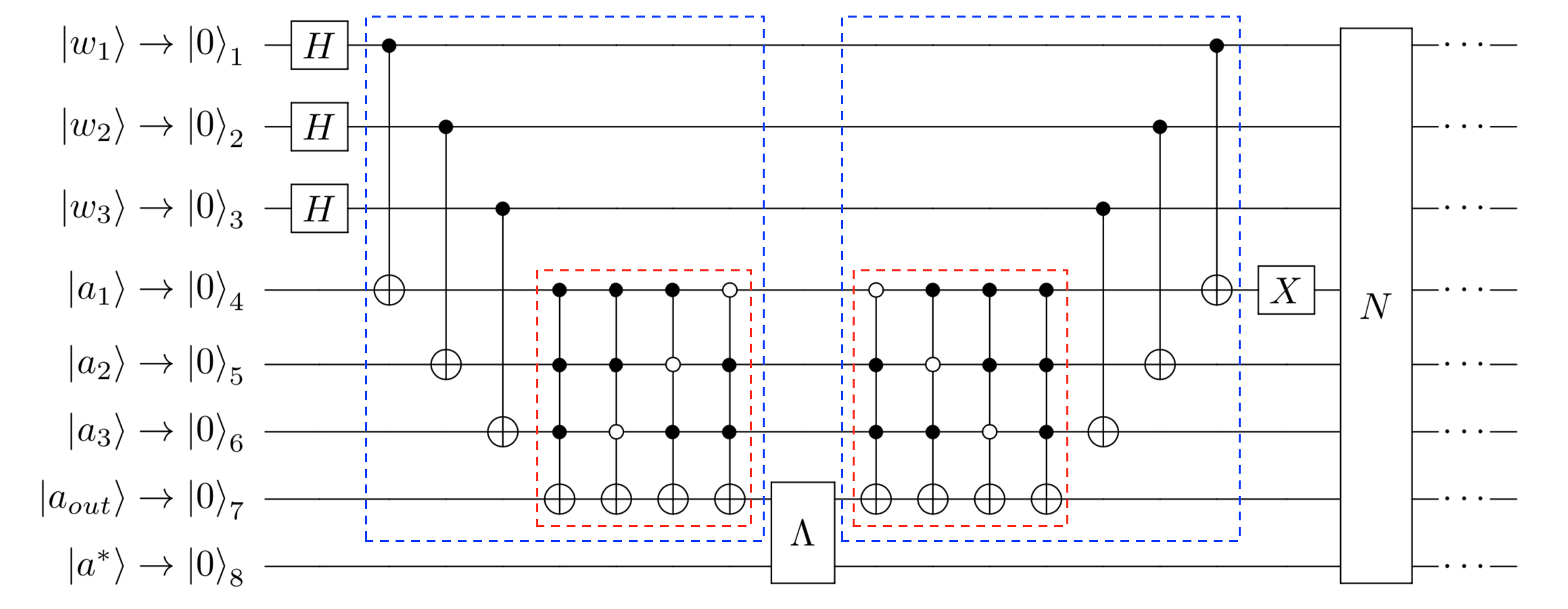}
\caption{Quantum circuit implementation of phase accumulation sub-routine for the example of a single neuron with 3 inputs and 3 weights.}
\label{fig:QCFON82}
\end{figure}

Next, we consider a single neuron with 3 weights and 3 inputs, as in figure \ref{fig:QCFON8}. As we will see below, this slight increase in inputs manifests itself in a significantly more complex gate implementation of the neuron action. Figure \ref{fig:QCFON82} shows the quantum circuit for the phase accumulation subroutine of this example.
\begin{figure}[h!]
\includegraphics[width=0.4\linewidth]{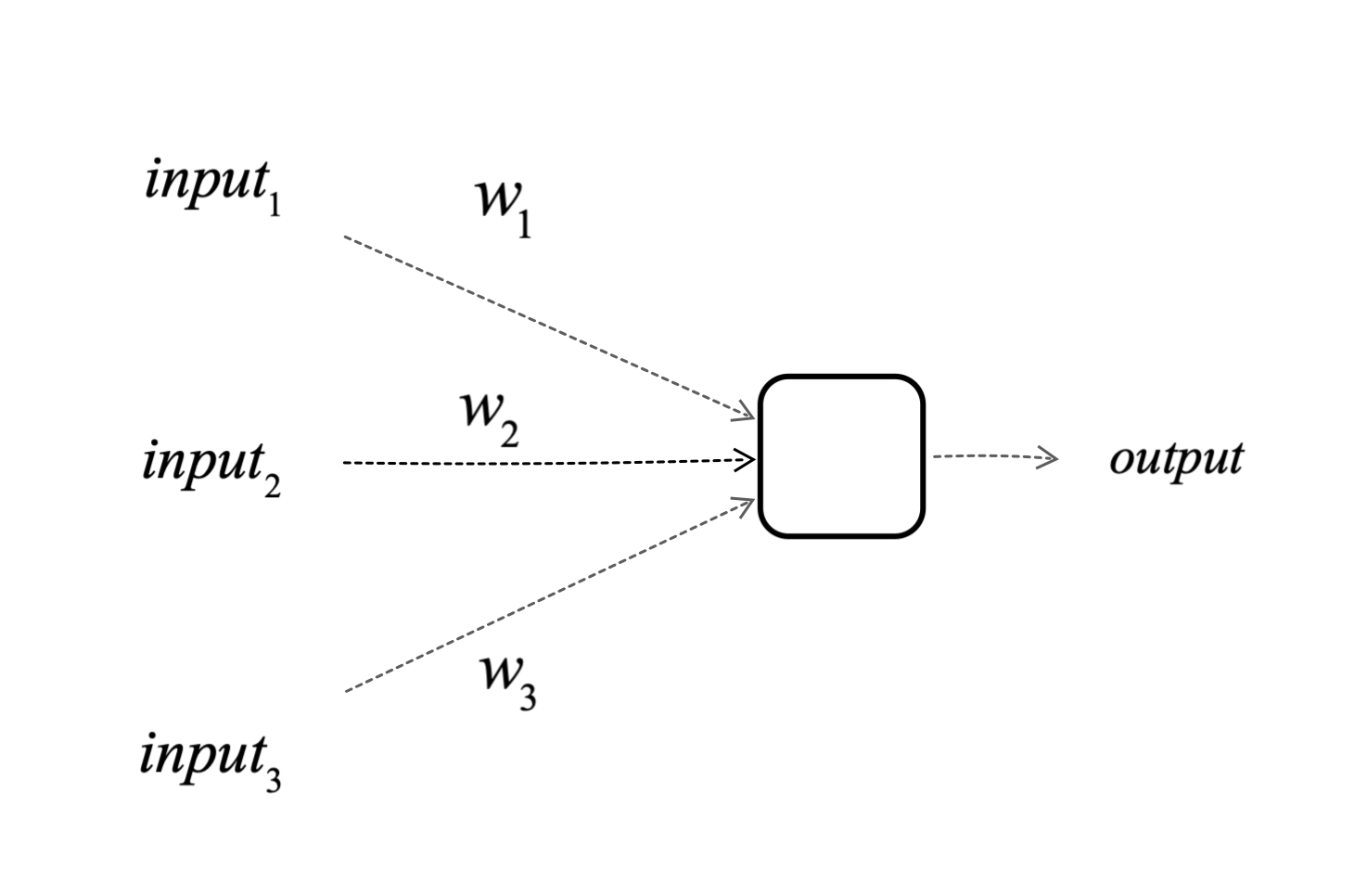}
\caption{Single neuron with 3 inputs and 3 weights.}
\label{fig:QCFON8}
\end{figure} 

As depicted in Figure \ref{fig:QCFON82} : qubits with index 1-3 take the role of the weights and qubits with index 4-6 are the three training inputs. Qubit 7 is an ancilla storing the output of the QBN. Qubit 8 carries the desired outcome. As before, the dashed blue box depicts the action of the QBN consisting of weighing the inputs with the weights, adding up the weighted input together with the subsequent activation fucntion (the gate implementation of the addition and activation is included in the dashed red box). After applying the oracle $\Lambda$, the action of the QBN is uncomputed (the circuit in the second dashed blue box). Finally, the full phase addition is repeated for $n$ times, where $n$ denotes again the amount of training input data. In this graph we only depict two data points $\{(a_1 , a_2, a_3, a^* )\}=\{(0,0,0,0),(1,0,0,0)\} $ of the training set and used ellipsis to indicate that the loop succeeds over all remaining training pairs. The two independent sets of full training data considered for this architecture are specified in Figure \ref{fig:qbnndata} below. \\

\begin{figure}[h]
\includegraphics[width=0.65\linewidth]{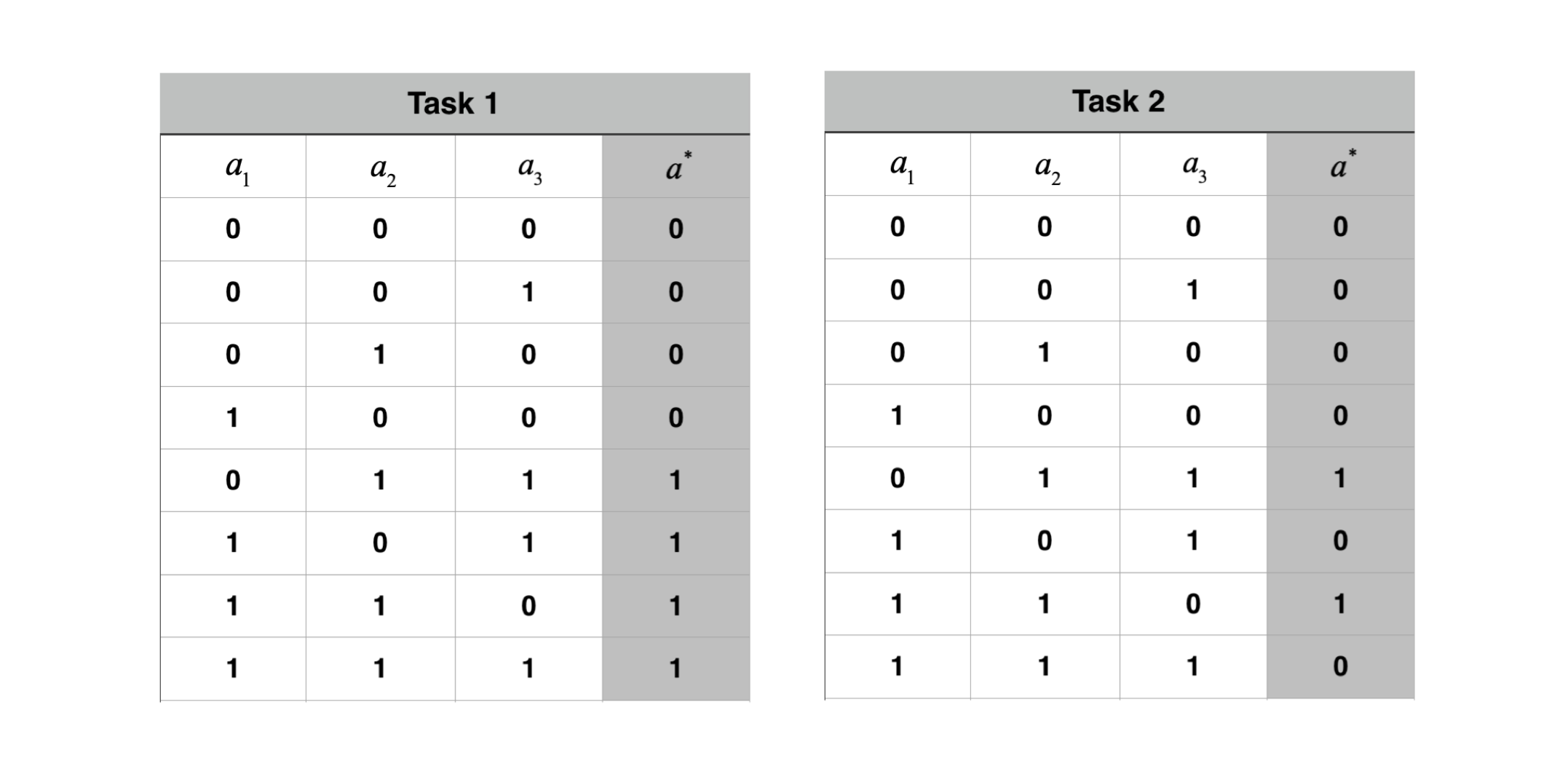}
\caption{\textit{Training data used for the example of a single neuron with 3 weights and 3 inputs:} each training tuple consists of three inputs $a_1$,$a_2$,$a_3$ and one desired output $a^*$. The aim of the training is to find weight configurations that can generate the desired output of the corresponding inputs. We consider two independent tasks (enumerated by 1 and 2) and study the training according to the data.}
\label{fig:qbnndata}
\end{figure} 

\begin{figure}[h!]
\includegraphics[width=0.95\linewidth]{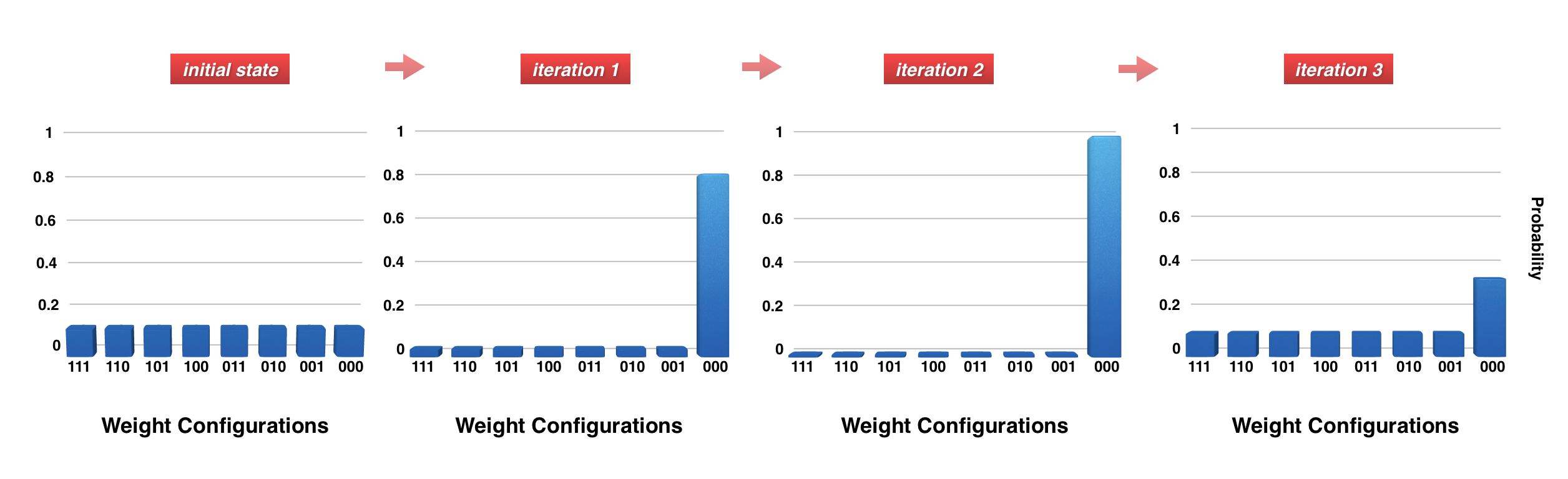}
\caption{\textit{Task 1, the probabilities of weight strings, evolving over the training cycles.} For Task 1 there exists only one optimal weight configuration, namely (000). The vertical bars present the probabilities of the weight configurations, for each iteration  of the full training cycle during the training. The probability of the optimal weight is amplified during the iterations and reaches its maximum after two iterations, in accordance to the optimal stopping time $k^*=\sqrt{8}\pi/4 \approx 2.22$ of standard Grover search.}
\label{fig:qbnn31data}
\end{figure}

\begin{figure}[h!]
\includegraphics[width=0.75\linewidth]{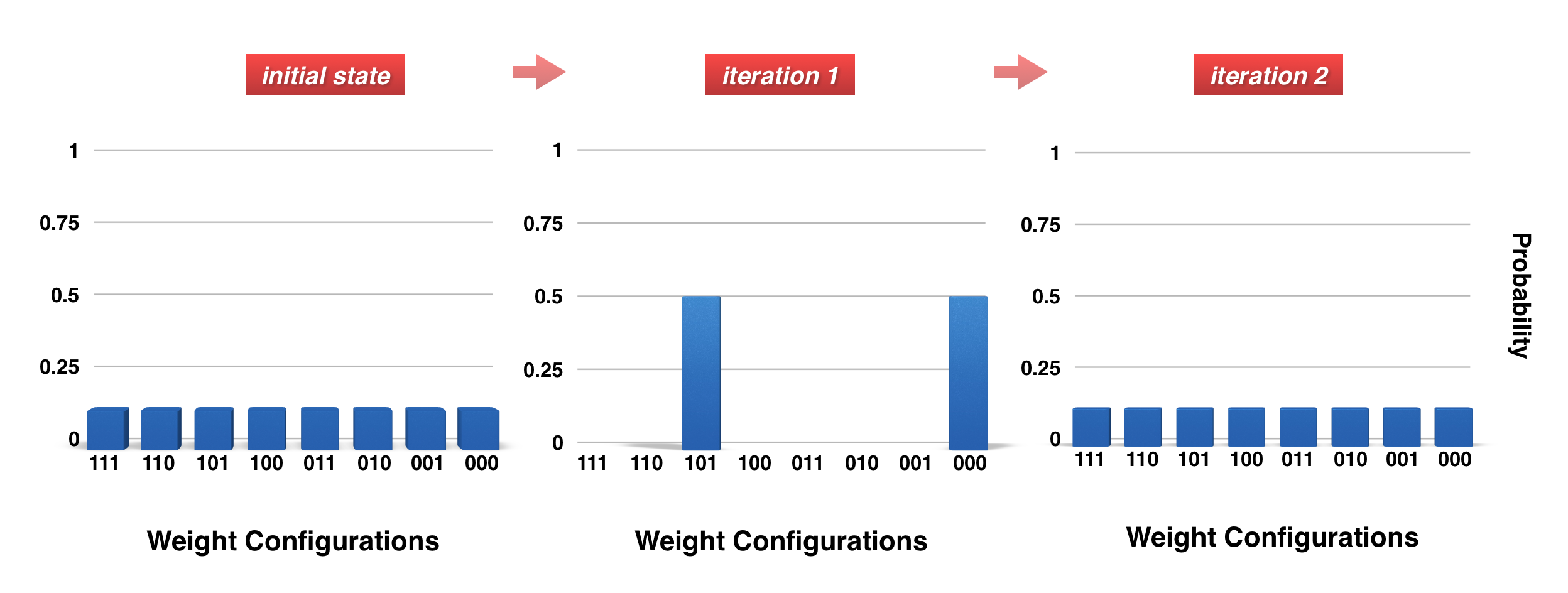}
\caption{\textit{Task 2, the probabilities of weight strings, evolving over the training cycles.} For Task 2 there are two optimal weight configurations, namely (000) and (101). The probabilities of the optimal weights are amplified during the iterations and reaches its maximum after the first iteration, in accordance to the optimal stopping time $k^*=\sqrt{8/2}\pi/4 \approx 1.57$ of Grover search with two possible solutions.}
\label{fig:qbnn31data2}
\end{figure}

During the loops of the full training (see Figure \ref{fig:PEfull} in the main text), the probabilities of the weight strings change. Here we present the probability evolution of the weight strings, in the quantum training with the highest $N_t$ for which there exists at least one optimal weight string: For task 1, Figure \ref{fig:qbnn31data} shows the probabilities of the weight strings evolving over the training cycles, and correspondingly figure \ref{fig:qbnn31data2} illustrates the evolution for task 2.

\subsection{3-layer 5-neuron network with 6 weights, 2 inputs and 1 ouput}
\begin{figure}[h!]
\includegraphics[width=0.5\linewidth]{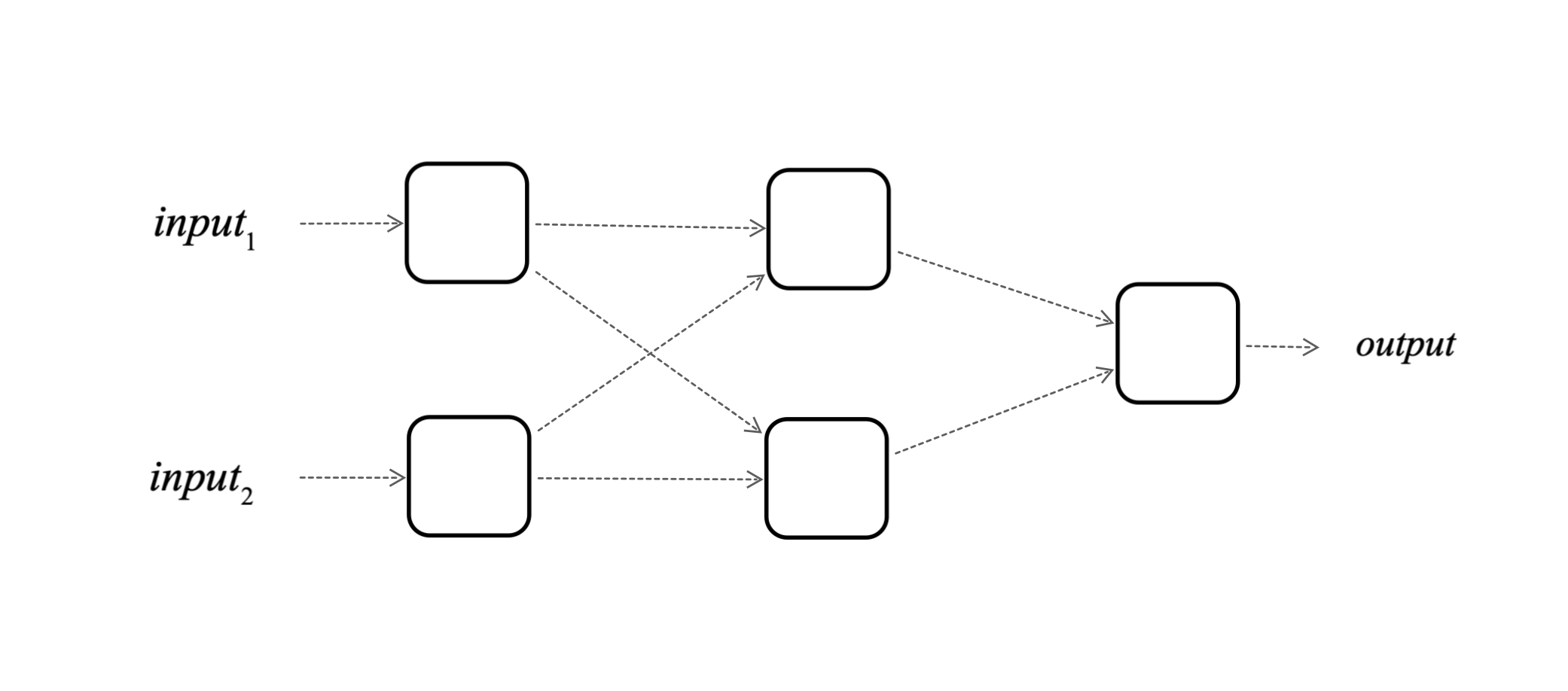}
\caption{Five-neuron network with layer configuration 2-2-1. The network takes two inputs and has in total 6 weights.}
\label{fig:qbnn221}
\end{figure}

\begin{figure}[h!]
\includegraphics[width=0.9\linewidth]{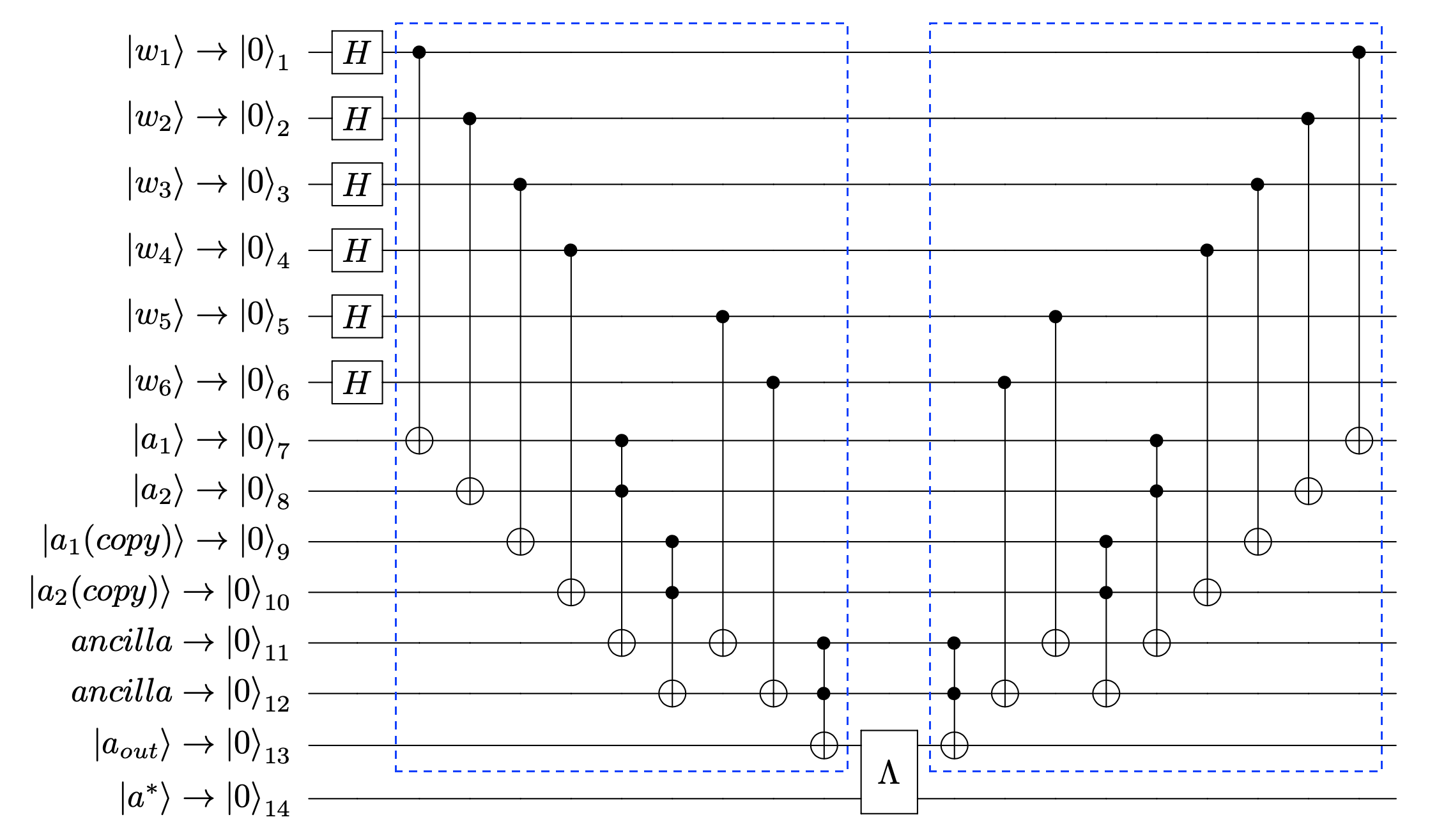}
\caption{\textit{Circuit implementation for a network with layer configuration 2-2-1, with 2 inputs and 6 weights.} The qubits indexed 1-4 are the weights for the first layer and qubits with index 5-6 are the weights for the second layer. Qubits with index 7-10 are 4 training inputs duplicated from 2 inputs in the first layer, for the fanout to the second layer.  (To save space in the diagram, the fan-outs in the first layer take place to the left, outside the diagram. Thus the inputs are already copied when they enter the depicted circuit.) Qubits 11-12 are ancillas storing the outputs within the QBNN. Qubit 13 encodes the desired outcome. The dashed blue box depicts the action of the QBNN. After applying the oracle $\Lambda$, the action of the QBNN is uncomputed. The circuit depicts one round of marking for one data point in the training set, for the phase accumulation process.}
\label{fig:qbnn2212}
\end{figure}

Analogously to before, we study the circuit implementation of the phase accumulation subroutine for a network with five neurons and layer configuration 2-2-1. This is depicted in figure \ref{fig:qbnn221} and \ref{fig:qbnn2212}. In total, there are two inputs and six weight states $|w_1\>$, $|w_2\>$, $\dots$, $|w_6\>$, leading to $2^6=64$ possible weight strings $|w_1 w_2 \dots w_6\>$. The two dashed blue boxes show the circuit for the computation and uncomputation of the neural network respectively.

\begin{figure}[h!]
\includegraphics[width=1\linewidth]{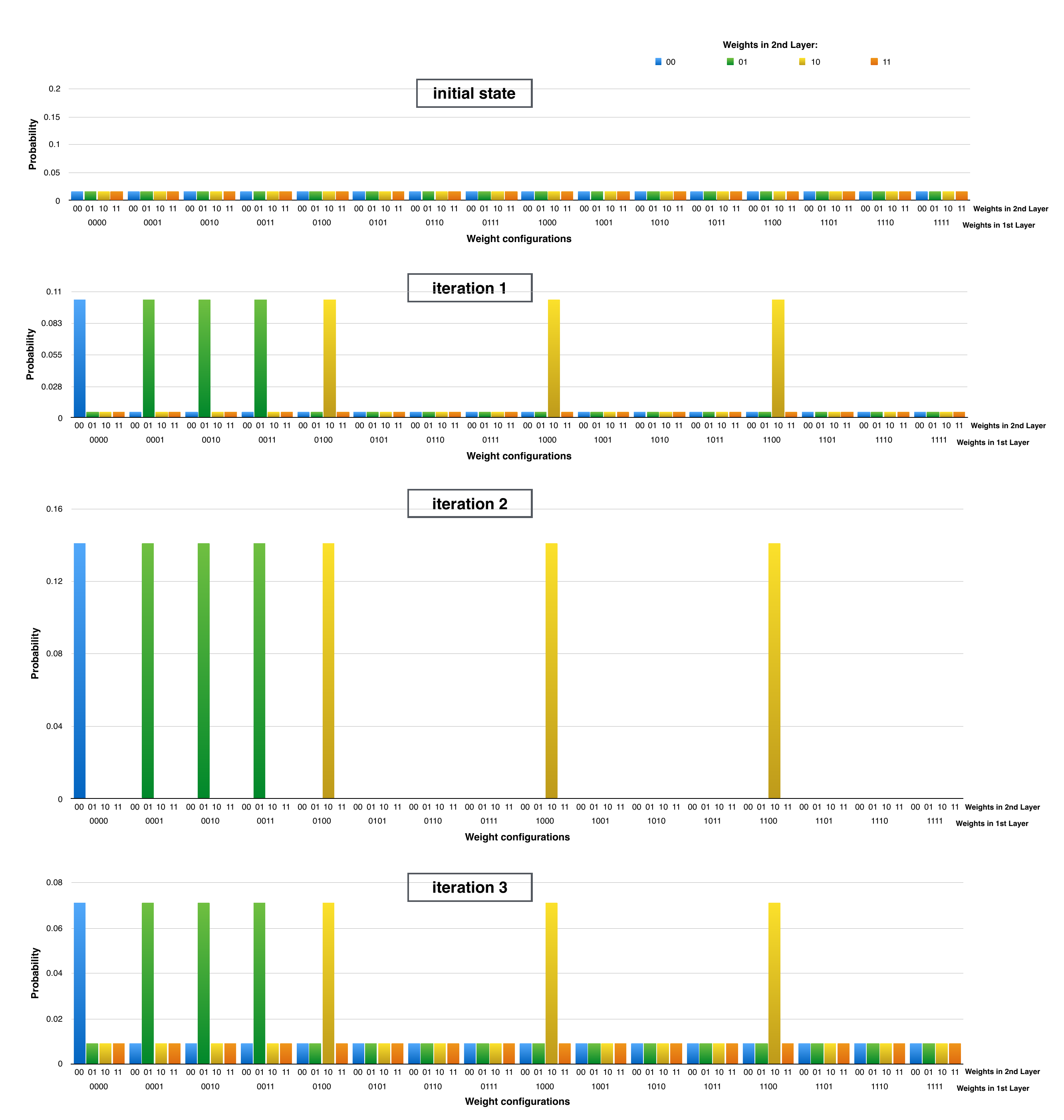}
\caption{\textit{The probability evolution of the weight strings in the 5-neuron QBNN with layer configuration 2-2-1, with 2 inputs and 6 weights.} For the training set $\{(a_1 , a_2, a^* )\}=\{(0,0,0),(1,0,0),(0,1,0),(1,1,1)\} $ there are 7 optimal weight configurations among the total $2^6=64$ configurations in the QBNN. In the Figure, the horizontal axis presents all the 64 weight configurations. The configurations which have the same weights the first layer are grouped together, with different colours representing the different weights $(00,01,10,11)$ in the second layer. The vertical bars presents the probabilities of the weight configurations, for each iteration during the training. The probabilities of the optimal weights are amplified over the iterations and reaches its largest amplification after two iterations. This result agrees with the expected stopping point given by $\sqrt{N/M} \pi/4$ to achieve the largest amplification, which is $\sqrt{64/7} \pi/4 \approx 2.37$ for this instance.
}
\label{fig:qbnn221data}
\end{figure}

For the QBNN in this Example, we use the training set  $\{(a_1 , a_2, a^* )\}=\{(0,0,0),(1,0,0),(0,1,0),(1,1,1)\} $). In figure \ref{fig:qbnn221data}, we present the probability evolution of the weight strings, in the quantum training with the highest $N_t$ for which there exists at least one optimal weight string.

\subsection{3-layer 6-neuron network with 8 weights, 3 inputs and 1 output}
\begin{figure}[h!]
\centering
\includegraphics[width=0.5\linewidth]{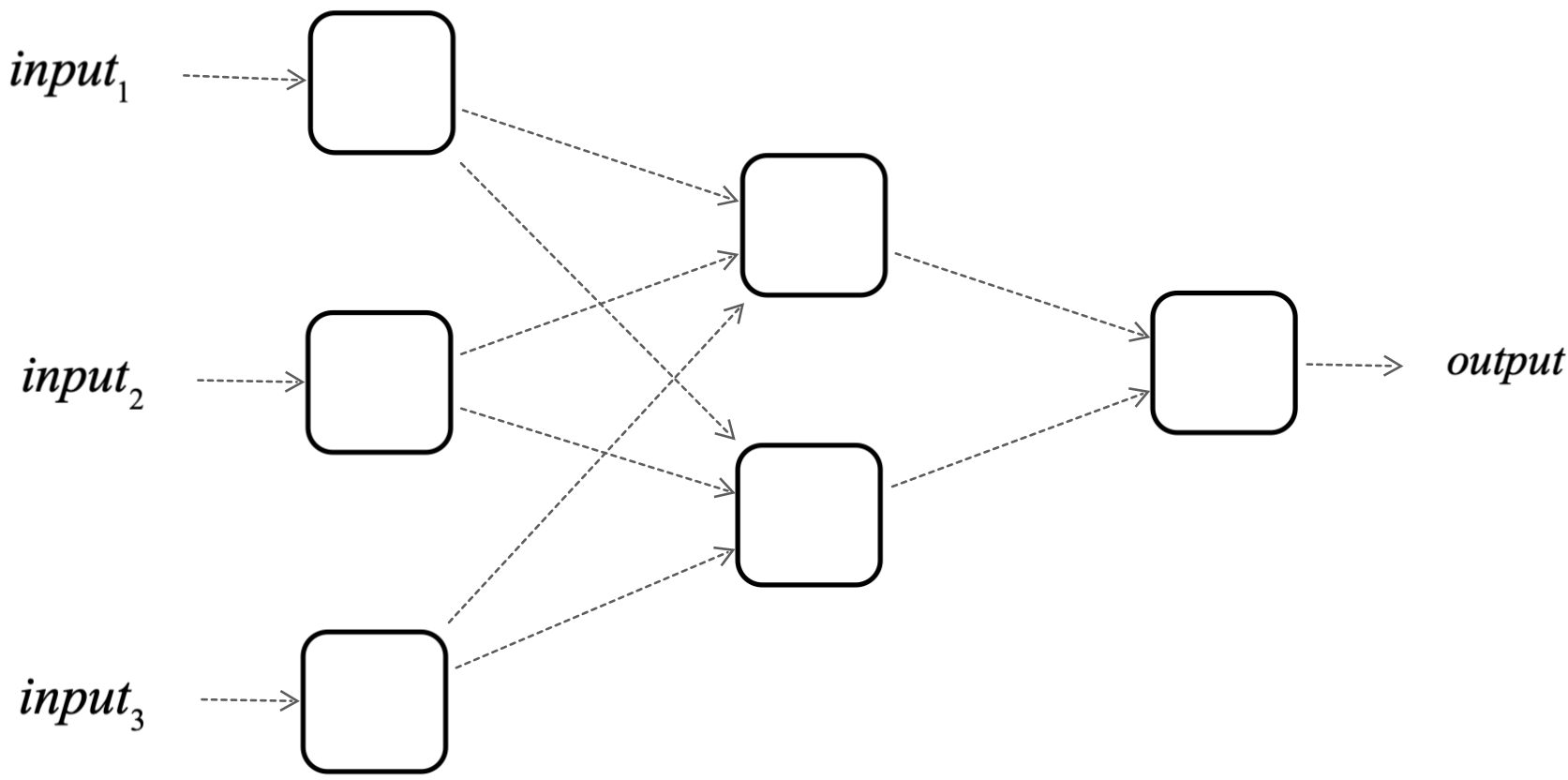}
\caption{Neural network with 6 neurons in layer configuration 3-2-1.}
\label{fig:classical321}
\end{figure}

\begin{figure}[h!]
\includegraphics[width=0.87\linewidth]{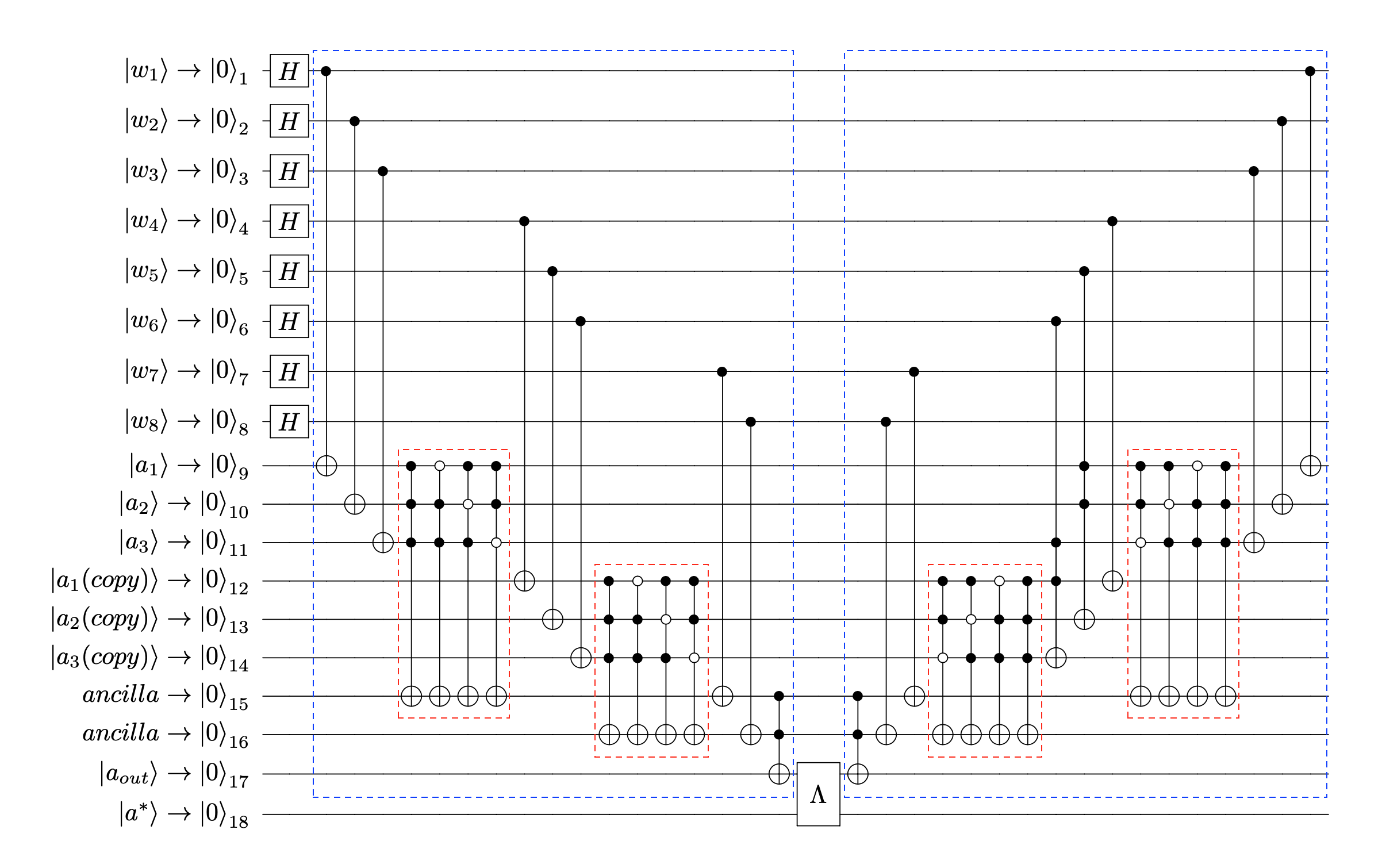}
\caption{\textit{Circuit implementation for a network with layer configuration 3-2-1.} Qubits 1-6 are the weights for the first layer and qubits 7-8 are the weights for the second layer. Qubits 9-14 are the 6 training inputs, duplicated from 3 inputs in the first layer, for the fanout to the second layer. (To save space in the diagram, the fan-outs in the first layer take place to the left, outside the diagram. Thus the inputs are already copied when they enter the depicted circuit.) Qubits with index 15-16 are ancillas storing the outputs of the neurons within the QBNN and qubit 17 encodes the output of the QBNN. Qubit 18 encodes the desired output. The dashed blue box contains the action of the QBNN, consisting of weighing the inputs with the weights, adding up the weighted input together with the subsequent activation fucntion (the gate implementation of the addition and activation is included in the dashed red box). After applying the oracle $\Lambda$, the action of the QBNN is uncomputed (the circuit in the second dashed blue box). The entire circuit depicts one round of marking for one data point in the training set, for the phase accumulation process.}
\label{fig:quantum321}
\end{figure}

We conclude the set of examples with a six-neuron network with layer configuration 3-2-1, with 3 inputs and 8 weights (see Figure \ref{fig:classical321}). This is one of the largest networks that can be implemented on quantum simulation platforms such as Huawei's computing cloud HiQ. The circuit implementation of a single cycle of phase accumulation is shown in figure \ref{fig:quantum321}. \\

\begin{figure}[h!]
\includegraphics[height=\dimexpr\pagegoal-\pagetotal-4\baselineskip\relax,
  width=\textwidth,
  keepaspectratio]{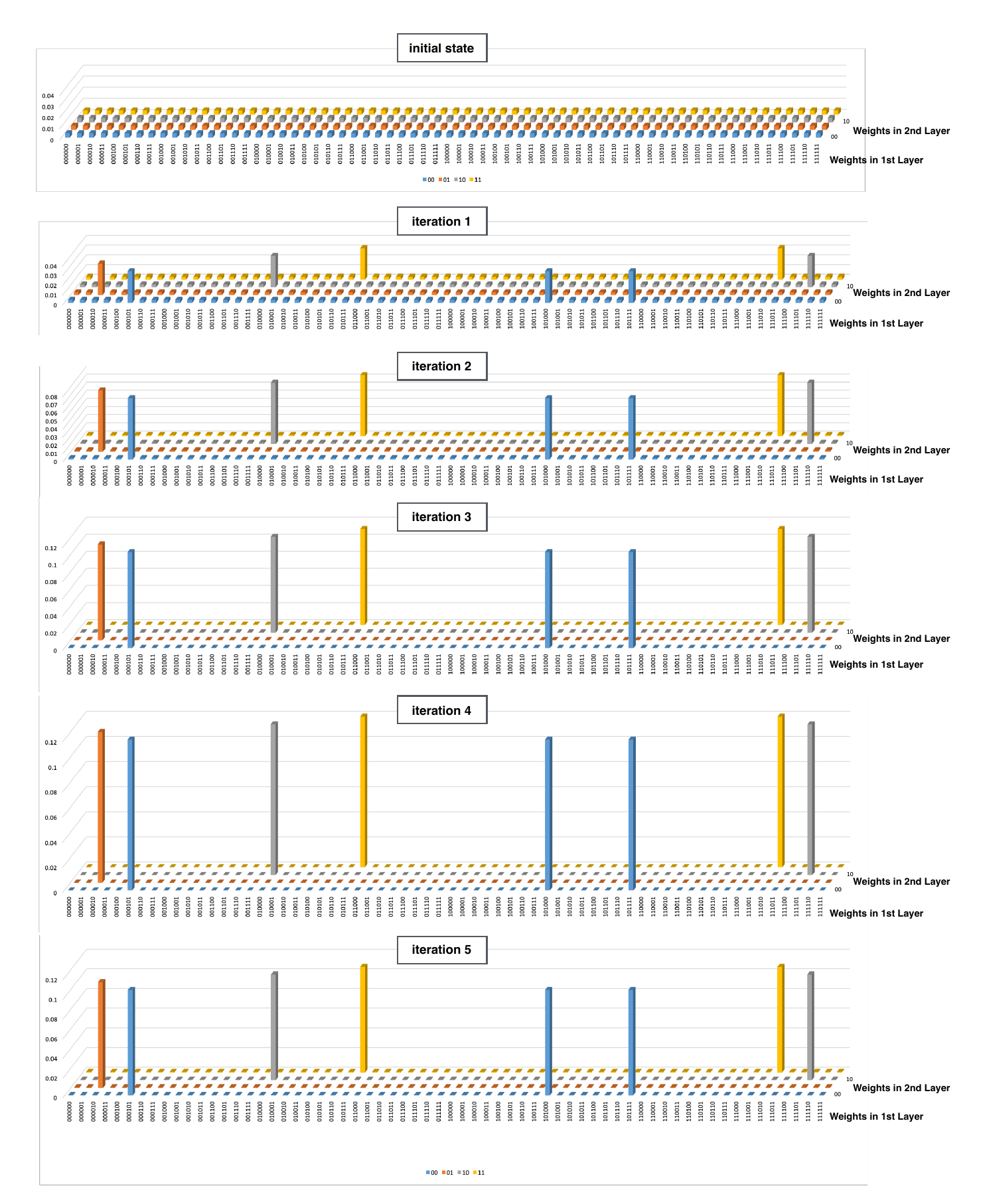}
\caption{\textit{Probabiliy evolution of the weight strings in the 6-neuron QBNN with configuration 3-2-1, task 2.} 
}
\label{fig:qbnn321data}
\end{figure}

We trained the QBNN in this example with the same two tasks presented in figure \ref{fig:qbnndata}. For instance of task 2, in figure \ref{fig:qbnn321data}, we present the probability evolution of the weight strings, in the quantum training with the highest $N_t$ for which there exists at least one optimal weight string. 

As in figure \ref{fig:qbnn321data}, for this example, there are 8 optimal weight configurations among the total $2^8=256$ configurations of the 8 binary weights in the QBNN. The horizontal plane presents all the 256 weight configurations. Each point in the grid represents one weight configuration, with two axes indicating the weights in the two layers respectively. The vertical bars present the probabilities of all the weight configurations, for each iteration during the training. The probability of the optimal weights are amplified over the iterations and reaches its top at iteration 4, in agreement with the expected stopping point given by $\sqrt{256/8} \pi  \approx 4.44$. \\

Figure \ref{fig:qbnn321data2} below shows the probability of obtaining an optimal weight string, as a function of the training cycle for both tasks.

\begin{figure}[h]
\includegraphics[width=0.8\linewidth]{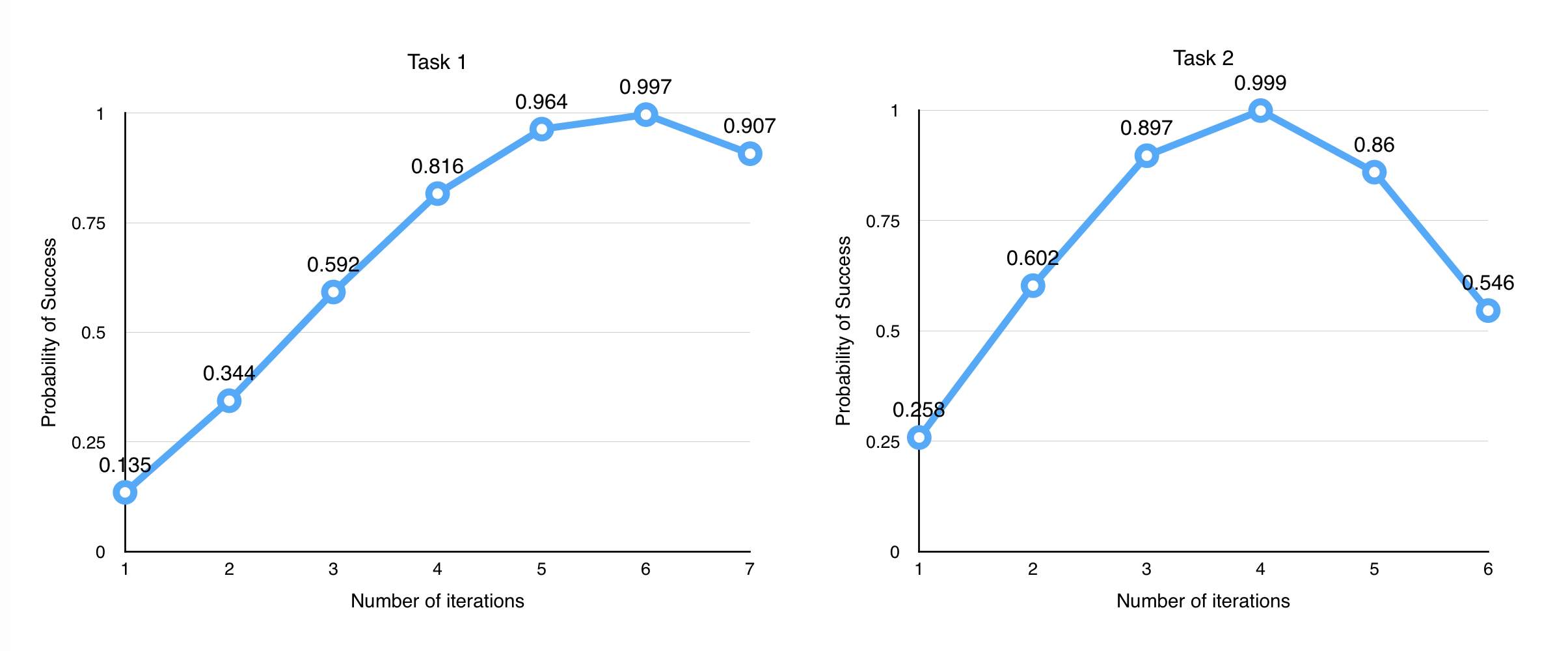}
\caption{\textit{Probability of obtaining an optimal weights, for a 6-neuron QBNN with layer configuration 3-2-1.} The graphs show the optimal stopping time of the training, for both task 1 and 2. We see that the probability of success is close to unity after 6 iterations for task 1 and 4 iterations for task 2. }
\label{fig:qbnn321data2}
\end{figure}

Recall the comparison between our quantum training and classical training stated in section \ref{perfPE}: classical brute force search calls the comparatory oracle ${N_C}^{classical}=n\times\tilde{N}$, whereas our quantum training algorithm only calls approximately $ N_C=N_G\times 2n\times (2n-1)$. 
Inserting the value of $n=8$ and $\tilde{N}=256$ for this example, one can see that even for this small network there is a quantum advantage in $N_C$: 
${N_C}^{classical}=8\times256$,  
whereas our quantum $N_C=8\times180$ (Task 1). Note that for large networks the quadratic advantage in 
$\tilde{N}$ will be much more significant, as discussed in the performance section. \\

\textit{*The implementations in this appendix are done via Huawei's Quantum Computing Cloud Platform HiQ (\url{https://hiq.huaweicloud.com}) and the open-source version code is available here: \url{https://github.com/phylyd/QBNN}}

\section{Equality for the number of qubits needed in the QBNN \label{app:noofqubits}}
In Section \ref{perfPE} we made use of the equality
\begin{align}
Q_{\text{input}} + Q_{\text{ancilla}} = Q_{\text{weight}} + Q_{\text{output}} \ , \label{app:bilance}
\end{align}
where $Q_{\text{input}}$ was the amount of input qubits, $Q_{\text{ancilla}}$ the amount of ancilla qubits, $Q_{\text{weight}}$ the amount of weight qubits and $Q_{\text{output}}$ the number of output qubits. Here, we derive this equality illustrated by two examples. \\

Let us begin with the elementary case of a single neuron, depicted in Figure \ref{qcircuit1} below. It is straightforward to count and verify the equality in Equation \ref{app:bilance}, by noting that $Q_{\text{ancilla}}=Q_{\text{output}}=1$ and that every input comes on an edge with an assigned weight, i.e. $Q_{\text{input}}=Q_{\text{weight}}$. 
Thus Equality \ref{app:bilance} holds for the case of a single neuron.\\

\begin{figure}
\[\Qcircuit @C=1em @R=.7em {
    \lstick{\ket{a_1}} & \multigate{4}{U} & \qw \\
    \lstick{\ket{w_1}} & \ghost{U} & \qw \\
    \lstick{\ket{a_2}} & \ghost{U} & \qw \\
    \lstick{\ket{w_2}} & \ghost{U} & \qw \\
    \lstick{\ket{ancilla}} & \ghost{U} & \rstick{\ket{output}} \qw 
}\]
\caption{\textit{the case of a single neuron} 
}
\label{qcircuit1}
\end{figure}
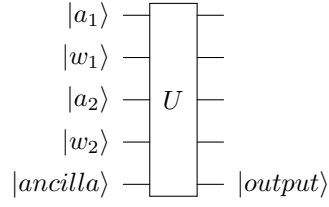

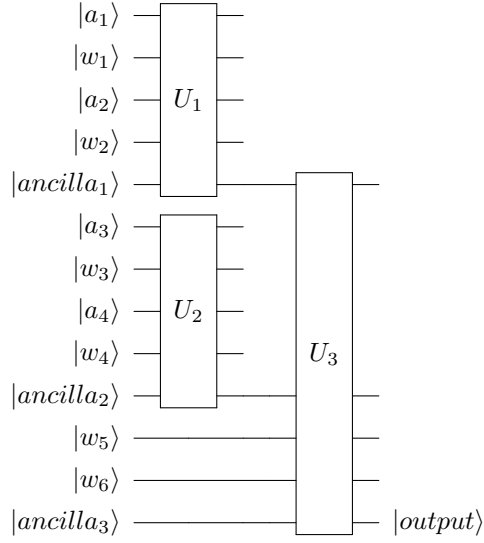
\begin{figure}
    \[ \Qcircuit @C=1em @R=.7em {
    \lstick{\ket{a_1}} & \multigate{4}{U_1} & \qw & & & \\
    \lstick{\ket{w_1}} & \ghost{U_1} & \qw & & & \\
    \lstick{\ket{a_2}} & \ghost{U_1} & \qw & & & \\
    \lstick{\ket{w_2}} & \ghost{U_1} & \qw & & & \\
    \lstick{\ket{ancilla_1}} & \ghost{U_1} &  \qw & \qw & \multigate{8}{U_3}& \qw \\
    \lstick{\ket{a_3}} & \multigate{4}{U_2} & \qw &   &   & \\
    \lstick{\ket{w_3}} & \ghost{U_2} & \qw &  &   & \\
    \lstick{\ket{a_4}} & \ghost{U_2} & \qw &  &   &  \\
    \lstick{\ket{w_4}} & \ghost{U_2} & \qw &  &   &  \\
    \lstick{\ket{ancilla_2}} & \ghost{U_2} & \qw & \qw & \ghost{U_3} & \qw \\
    \lstick{\ket{w_5}} & \qw & \qw & \qw & \ghost{U_3} & \qw \\
    \lstick{\ket{w_6}} & \qw & \qw & \qw & \ghost{U_3} & \qw \\
    \lstick{\ket{ancilla_3}} & \qw & \qw & \qw & \ghost{U_3} & \rstick{\ket{output}} \qw \\
}\]
    \caption{\textit{the case of 3 neuron in 2 layers} }
    \label{qcircuit2}
\end{figure}

Next, we consider a more involved network with three neurons in two layers, see Figure \ref{qcircuit2}.\\

Direct counting verifies that Equation \ref{app:bilance} also holds here. The reason in general is that in the first layer the input qubits and weight qubits always come in pairs, i.e.
\begin{align}
Q_{\text{input}}= Q_{\text{weight}}^{(in)} \ ,
\end{align}
where the superscript denotes that the count is for the weights in the first layer. \\

In the first layer and hidden layers, the ancilla qubits from the previous layer carry the output of the neuron computation, together with fan-out ancilla qubits, serve as inputs for the next layer, and each of them is paired with a weight qubit in the hidden layers. Mathematically, this condition reads
 \begin{align}
Q_{\text{ancilla}}^{(in)}+Q_{\text{ancilla}}^{(hidden)} = Q_{\text{weight}}^{(hidden)} \ ,
\end{align}
where $hidden$ denotes that the equality holds for any hidden layer between the input and output layer of the network. For the output layer each output qubit corresponds to an ancilla qubit in this layer. Hence,
 \begin{align}
Q_{\text{ancilla}}^{(out)} = Q_{\text{output}} \ .
\end{align}
Altogether, we see that the equality
\begin{align}
Q_{\text{input}} + Q_{\text{ancilla}} = Q_{\text{weight}} + Q_{\text{output}} 
\end{align}
follows, for general QBNN according to our design.

\section{Correlation between $n$ and $N$ for a two hidden layer feedfoward network \label{2hidden}}

In \cite{huang2003learning} the author proved that for a two-hidden-layer feedforward network with m output neurons, the number of hidden nodes that are enough to learn N samples with negligibly small error is given by:
 \begin{equation}
 2\sqrt{(m+2)N}
 \end{equation}
 Specifically, the sufficient number of hidden nodes in the first
layer is suggested to be
\begin{equation}
L_1 = \sqrt{(m+2)N}+2 \sqrt{N/(m+2)}
 \end{equation}

and in the second layer is suggested to be
\begin{equation}
L_2 = m\sqrt{N/(m+2)}
 \end{equation}

In this architecture, the number of weights between the two hidden layers is the product of the number of nodes in the two hidden layers:

\begin{align}
 L_1 \times  L_2 \\ =({\sqrt{(m+2)N}+2 \sqrt{N/(m+2)}})\times( m\sqrt{N/(m+2)})\\ =
 mN+2Nm/(m+2)
  \end{align}
 

 
The total number of weights in this optimal architecture $N_{total}$ is larger than $ L_1 \times  L_2 $ (Since we also have weights between input layer and first hidden layer, weights between second hidden layer and output layer). Thus:

 \begin{equation}
 N_{total} > L_1 \times  L_2 > mN
 \end{equation}
 
 Using the notation we use in our paper, i.e.$ N_{total} \mapsto N, N \mapsto n $ we have :
 
  \begin{equation}
 N  > m \times n
 \end{equation}
 
 where $N$ is the number of weight qubits and n is the number of training samples, $m$ is the number of neurons in the output layer. 
 Since $m$ is an integer larger than 1, so $ N > mn > n $. Now we have derived the correlation $ n < N$ used in the maintext.

\end{document}